\documentclass[pra,showpacs,floatfix,superscriptaddress,twocolumn]{revtex4}

\pdfoutput=1

\usepackage{graphics,graphicx}
\usepackage{mathrsfs}
\usepackage{amsmath,amsfonts,amssymb}
\usepackage{subfigure}
\usepackage{color}

\begin{document}

\title{Characterization of the low temperature properties of a
  simplified protein model} 

\author{Johannes-Geert Hagmann}
\affiliation{Laboratoire de Physique, Ecole
Normale Sup\'erieure de Lyon, CNRS, 46 All\'ee d'Italie, 69364 Lyon,
France} 
\author{Naoko Nakagawa}
\affiliation{Department of Mathematical Sciences, Ibaraki University,
Mito, Ibaraki 310-8512, Japan} 
\author{Michel Peyrard}
\affiliation{Laboratoire de Physique, Ecole
Normale Sup\'erieure de Lyon, CNRS, 46 All\'e d'Italie, 69364 Lyon,
France}

\date{\today}

\begin{abstract}
Prompted by results that showed that a simple protein model, the
frustrated G\=o model, appears to exhibit a transition reminiscent of
the protein dynamical transition, we examine the validity of this model
to describe the low-temperature properties of proteins.  First, we
examine equilibrium fluctuations.  We calculate its incoherent
neutron-scattering structure factor and show that it can be well
described by a theory using the one-phonon approximation.  By performing
an inherent structure analysis, we assess the transitions among energy
states at low temperatures.  Then, we examine non-equilibrium
fluctuations after a sudden cooling of the protein.  We investigate the
violation of the fluctuation--dissipation theorem in order to analyze
the protein glass transition.  We find that the effective temperature
of the quenched protein 
deviates from the temperature of the thermostat, however it relaxes towards
the actual temperature with an Arrhenius behavior as the waiting time
increases.  These results of the equilibrium and non-equilibrium studies
converge to the conclusion that the apparent dynamical transition of
this coarse-grained model cannot be attributed to a glassy behavior.
\end{abstract}


\maketitle

\section{Introduction}

Proteins are fascinating molecules due to their ability to play many
roles in biological systems. Their functions often involve complex
configurational changes. Therefore the familiar aphorism that ``form is
function'' 
should rather be replaced by a view of the ``dynamic personalities of
proteins''\cite{Henzler}. This is why proteins are also intriguing for
theoreticians because they provide a variety of yet unsolved questions.
Besides the dynamics of protein folding, the rise in the time averaged
mean square fluctuation $\langle \Delta r^2\rangle$ occurring at
temperatures around $\approx 200K$, sometimes called the ``protein dynamic
transition'' \cite{doster,parak,frauenfelder1}
is arguably the most considerable candidate in the search of unifying
principles in protein dynamics. Protein studies lead to the concept of {\em
energy landscape} \cite{Frauenfelder2,Bryngelson}. According to  
this viewpoint a protein is a system which explores a complex landscape
in a highly multidimensional space and some of its properties can be
related to an incomplete exploration of the phase space. The protein
glass transition, in which the protein appears to ``freeze'' when it is
cooled down to about $200\;$K is among them. Protein folding too can be
related to this energy landscape. The famous kinetic limitation known as
the Levinthal paradox, associated to the difficulty to find the native
state among a huge number of possible configurations, is partly solved
by the concept of a funneled landscape which provides a bias towards the
native state.

\smallskip
These considerations suggest that the dynamics of the exploration of protein
phase space deserves investigation, particularly at low
temperature where the dynamic transition occurs. But, in spite of remarkable
experimental progress which allows to ``watch protein in action
in real time at atomic resolution'' \cite{Henzler}, experimental studies
at this level of detail are nevertheless extremely difficult. Further
understanding from models can help in analyzing the observations and
developing new concepts. However, studies involving computer modeling to
study the dynamics of protein fluctuations are not trivial either
because the range of time scales involved is very large. This is why
many meso-scale models, which describe the protein at scales that are 
larger than
the atom, have been proposed. Yet, their validity to adequately describe
the qualitative features of a real protein glass remains to be tested.

\smallskip
In this paper we examine a model with an
intermediate level of complexity. This frustrated G\=o model
\cite{Clementi,Karanicolas} is an off-lattice model showing
fluctuations at a large range of time scales. It is though
simple enough to allow the investigation of time scales which can be up
to $10^9$ times larger than the time scales of small amplitude
vibrations at the atomic level. The model, which includes a slight
frustration in the dihedral 
angle potential which does not assume a minimum for the positions of the
experimentally determined structure, exhibits a much richer behavior
than a standard G\=o model. Besides folding one observes 
a rise of fluctuations above a specific temperature, analogous to
a  dynamical transition \cite{nakagawa2,nakagawa1},
and the coexistence of two folded states. 
This model has been widely used and
it is therefore important to assess to what extent it can describe
the qualitative features of protein dynamics 
beyond the analysis of folding for which it was originally designed. 
This is why we focus our attention on its
low-temperature properties in an attempt to determine if a fairly simple
model can provide some insight on the protein dynamical
transition. The purpose of the present article is to clarify the origin of the
transition in the computer model, and to determine similarities and
differences with respect to experimental observations.
Although the calculations are performed with a specific model, the
methods are more general and even raise some questions for experiments,
especially concerning the non-equilibrium properties.

\bigskip
This article is organized as follows. The numerical findings relating to
the ''dynamical transition''from previous studies
\cite{nakagawa2,nakagawa1} are presented in Sec.~\ref{sec:model}. 
As a very large body of
experimental studies of protein dynamics emanates from neutron
scattering experiments, it 
is rational to seek a connection between theory and experiment by
studying the most relevant experimental observable for dynamics, the
incoherent structure factor (ISF). We calculate the ISF from molecular dynamics
simulations of the model in Sec.~\ref{sec:strfact}. We show that its 
main features can be well reproduced by a theoretical analysis based
on the one-phonon approximation, which indicates that, at low
temperature, the dynamics of the protein within this model
takes place in a single minimum
of the energy landscape. 
Sec.~\ref{sec:is} proceeds to an inherent
structure analysis to examine how the transitions among energy states  
start to play a role when
temperature increases. As the freezing of the protein dynamics at low
temperature is often called a ``glass transition'', this raises the
question of the properties of the model protein in non-equilibrium
situations. In Sec.~\ref{sec:fdt} we examine the violation of
the fluctuation--dissipation theorem after a sudden cooling of the
protein.
We find that the effective temperature of the quenched protein, deduced
from the Fluctuation-Dissipation Theorem (FDT) deviates from the
temperature of the thermostat, however it relaxes towards the actual
temperature with an an Arrhenius behavior as the waiting time increases.
This would imply that the dynamics of the protein model is very slow but
not actually glassy.  This method could be useful to distinguish very
slow dynamics from glassy dynamics, in experimental cases as well as in
molecular dynamics simulations.  Finally Sec.~\ref{sec:discussion}
summarizes and discusses our results.

\section{A dynamical transition in a simple protein model?}
\label{sec:model}

Following earlier studies \cite{nakagawa2,nakagawa1,hagmann1} we chose
to study a small protein containing the most common types of secondary structure
elements ($\alpha\; $helix, $\beta\; $sheets and loops), protein $G$, the
$B_1$ domain of immunoglobulin binding protein \cite{clore} (Protein
Data Bank code 2GB1). It contains 56 residues, with one $\alpha\;$helix
and four $\beta\; $strands forming a $\beta\;$sheet.
We describe it by an off-lattice G\=o model with a slight frustration
which represents its geometry
in terms of a single particle per residue, centered at the location of
each $C_{\alpha}$ carbon in the experimentally determined tertiary structure. 
The interactions between these residues
do not distinguish between the type of amino acids. Details on the simulation
  process and the parametrization of the model are presented in the Appendix.
In spite of its simplicity, this model
appears to exhibit properties which are reminiscent of the protein
dynamical transition. This shows up when one examines the temperature
dependence of its mean-squared fluctuations \cite{nakagawa2} by calculating
the variance  $\Delta r^2$ 
of the residue distances to the center of mass as a function of
temperature, defined by 
\begin{eqnarray}
\Delta r^2 &=& \frac{1}{N}\sum_{i=1}^{N}\left( \overline{r_{i0}^2} -
  \overline{r_{i0}}^2 \right) \ \ \ . 
\end{eqnarray}
Here, $N$ denotes the number of residues, and $r_{i0}$ is distance of
residue $i$ with respect to the instantaneous center of of mass. The
average $\overline{A}$ of the observable $A(t)$ is the time average $ 
\overline{A}=\frac{1}{T}\int_{0}^{T}dt\ A(t)$.  The variances of 20
trajectories (Langevin dynamics simulations, each $3\cdot 10^{7}$ time
units long) were averaged for each temperature point ($\langle
\cdot\rangle$ denotes the average over independent initial
conditions). 
\begin{figure}[h]
\centering
\includegraphics[width=7.6cm]{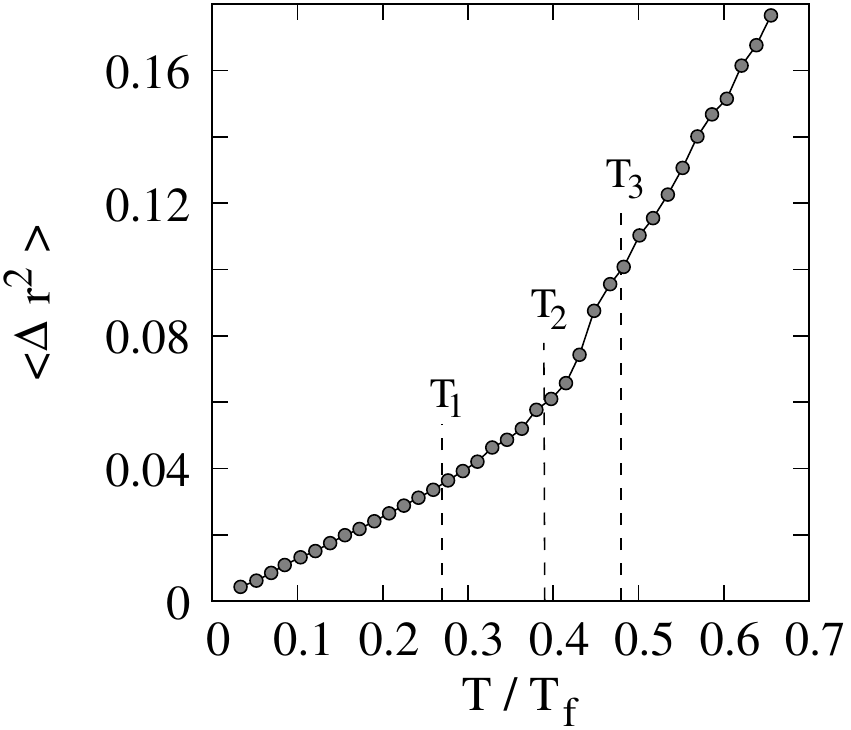}
\caption{Average mean-squared distance fluctuations
  $\langle \Delta r^2 \rangle$ as a function of temperature for protein
  G. Data adapted from
  \cite{nakagawa2}. The temperatures marked $T_1$, $T_2$, $T_3$ are the
  temperatures studied in Sec.~\ref{sec:is}.}
\label{fig:deltar2}
\end{figure}
\noindent 
Figure \ref{fig:deltar2}  shows the evolution of $\langle
\Delta r^2 \rangle$ as a function of temperature.  It exhibits
a crossover in the
fluctuations in the temperature range $T/T_f = [0.4,0.5]$ resembling the
transition observed for hydrated proteins in neutron scattering and
M\"ossbauer spectroscopy experiments \cite{doster,parak,frauenfelder1}, the
so-called \textit{dynamical transition}. Above $T/T_f = 0.4$, the
fluctuations increase quickly with temperature whereas a smaller, linear,
growth is observed below. One may wonder whether the complexity of
the protein structure, reflected by the G\=o model, is sufficient to
lead to a dynamical transition or whether, 
notwithstanding the resemblance of the
onset of fluctuations in the present model and the experimentally
determined transition, different physical and not necessarily related
events may contribute to the curves which by coincidence look
similar. One can already note that, for a folding temperature in
the range $330-350K$, the range $0.4-0.5\ T/T_f$ corresponds to
$132-175K$, lower than the experimentally observed transition
occurring around $180-200K$.  If it had been confirmed the
observation of a dynamical transition in a fairly simple protein model
would have been very useful to shine a new light on this transition which
is still not fully understood.
It is generally agreed that it is hydration-dependent
\cite{frauenfelder}, but still different directions for a microscopic
interpretation are being pursued, suggesting the existence \cite{chen}
or non-existence \cite{sokolov1} of a transition in the solvent
coinciding at the dynamical transition temperature. Recently, a
completely different mechanism based on percolation theory for the
hydration layer has been proposed \cite{kataoka1}. The precise nature of
the interaction between the solvent and proteins, and the driving factor
behind the transition, hence still remain to be understood. The
dynamical transition has often been called the \textit{protein glass
  transition} due its similarity with some physical properties of
structural glasses at low temperatures. In particular, it was pointed
out that, for both glasses and protein solutions, the transition goes
along with a crossover towards non-exponential relaxation rates at low
temperatures. The comparison is however vague since the glass transition
itself and notably its mechanism are ongoing subjects of research and
debate.

Our goal in this paper is to clarify the
origin of the numerically observed transition, which moreover gives hints
on the possibilities and limits of protein computer models.

\section{Analysis of the incoherent structure factor}
\label{sec:strfact}

If computer models of proteins are to be useful they must go beyond a
simple determination of the dynamics of the atoms, and make the link
with experimental observations. This is particularly important for the
``dynamical transition'' because its nature in a real protein is not
known at the level of the atomic trajectories. It is only
observed indirectly through the signals provided by
experiments. Therefore a valid analysis of the transition observed in the
computer model must examine it in the same context,
i.e.\ determine its consequences on the experimental observations.

Along with NMR and M\"ossbauer spectroscopy, neutron scattering methods
have been among the most versatile and valuable tools to provide insight
on the internal motion of proteins \cite{review1,smith91}.
Indeed, the thermal neutron wavelength being of the order
of \AA ngstr\"oms and the kinetic energy of the order of $meV$s,
neutrons provide an adequate probe matching the length- and frequency
scales of atomic motion in proteins.
An aspect brought forward in the discussion of the
dynamical transition in view of the properties of glassy materials is
the existence of a \textit{boson peak} at low frequencies in neutron
scattering spectra \cite{doster99}. Such a broad peak appears to be a
characteristic feature of unstructured materials as compared to the
spectra of crystals.

\subsection{Incoherent structure factor from molecular dynamics
  trajectories}

In neutron scattering the vibrational and conformational
changes in proteins appear as a quasielastic contribution
to the dynamic structure factor $S(\mathbf{q},\omega)$ which contains
crucial information about the dynamics on different time- and length
scales of the system. In scattering experiments one measures
the double-differential scattering cross section 
$d^2 \sigma/(d\Omega dE)$ which gives the probability of finding a neutron
in the solid angle element $d\Omega$ with an energy exchange $dE$ after
scattering. The total cross-section of the experiment is obtained by
integration over all angles and energies. Neglecting magnetic
interaction and only considering  the short-range nuclear forces, the
isotropic scattering is characterized by a single parameter $b_i$, the
scattering length of the atomic species $i$ \cite{bee}, which can be a
complex number with a non-vanishing imaginary part accounting for
absorption of the neutron. If one
defines  the average over different spin states $b^{coh}=\left|\langle b
  \rangle\right|$ as the coherent scattering length, and the root mean
square deviation  $b^{inc}=\sqrt{\langle |b|^2 \rangle -  \left|\langle
    b \rangle\right|^2}$ as the incoherent scattering length, the
double-differential cross section arising from the scattering of a
monochromatic beam of neutrons with incident wave vector $\mathbf{k}_0$
and final wave vector $\mathbf{k}$  by $N$ nuclei of the sample can be
expressed as \cite{bee} 
\begin{eqnarray}
\frac{d^2 \sigma}{d\Omega dE} &=& \frac{N}{\hbar}
\frac{|\mathbf{k}|}{|\mathbf{k}_0|} \left(b^{coh}\right)^2
S_{coh}(\mathbf{q},\omega) \nonumber \\ & &+
\frac{N}{\hbar}\frac{|\mathbf{k}|}{|\mathbf{k}_0|}
\left(b^{inc}\right)^2  S_{inc}(\mathbf{q},\omega) \ \ \ \ , 
\end{eqnarray}
where $\mathbf{q}=\mathbf{k}-\mathbf{k}_0$ is the wave vector transfer
in the scattering process and $\mathbf{r}_i$ denote the time-dependent
positions of the sample nuclei and 
the coherent and incoherent dynamical structure factors are 
\begin{align}
S_{coh}(\mathbf{q},\omega)&=\frac{1}{2\pi
  N}\sum_{i,j}\int_{-\infty}^{\infty}dt\ e^{-i\omega t} \langle
e^{-i\mathbf{q}(\mathbf{r}_i(t)-\mathbf{r}_j(0))} \rangle , \\ 
S_{inc}(\mathbf{q},\omega)&=\frac{1}{2\pi
  N}\sum_{i}\int_{-\infty}^{\infty}dt\ e^{-i\omega t} \langle
e^{-i\mathbf{q}(\mathbf{r}_i(t)-\mathbf{r}_i(0))} \rangle . 
\end{align}
The coherent structure factor contains contributions from the position
of all nuclei. The interference pattern of
$S_{coh}(\mathbf{q},\omega)$ contains the average (static) 
structural information on the
sample, whereas the incoherent
structure factor $S_{inc}(\mathbf{q},\omega)$ monitors the average of atomic
motions as it is mathematically equivalent to the Fourier transform in
space and time of the particle density autocorrelation function. In
experiments on biological samples, incoherent scattering from hydrogens
dominates the experimental spectra \cite{smith91} unless deuteration of
the molecule and/or solvent are used.

\smallskip
Since the
G\=o-model represents a reduced description of the protein and the
locations of the individual atoms in the residues are not resolved, we
use ''effective'' incoherent weights of equal value for the effective
particles of the model located in the position of the
$C_{\alpha}$-atoms. Such a coarse grained view assumes that the average
number of hydrogens atoms and their location in the residues is
homogeneous, which is of course a crude approximation in particular in
view of the extension and the motion of the side chains. These
approximations are nevertheless acceptable here as we do not intend to
provide a quantitative comparison with experimental results considering
the simplifications and the resulting limitations of the model.

We generated Langevin and Nos\'e-Hoover
dynamic trajectories of length $t=10^5$ time units, i.e.\ about
1000 periods of the slowest vibrational mode of the protein, after an
equilibration of equal length for protein G at temperatures in the
interval $T/T_f = [0.0459,0.9633]$. 
To compute the incoherent structure factor for the G\=o-model of
protein G, we use \textit{nMoldyn} \cite{kneller} to analyze the
molecular dynamics trajectories generated at different temperatures. The
data are spatially averaged over $N_q=50$ wave vectors sampling
spheres of fixed modules $|\mathbf{q}|=2,3,4$ \r{A}$^{-1}$, and the
Fourier transformation is smoothed by a Gaussian window of width
$\sigma=5\%$ of the full length of the trajectory. Prior to the
analysis, a root-mean square displacement alignment of the trajectory
onto the reference structure at time $t=0$ is performed using virtual
molecular dynamics (VMD)
\cite{vmd}. Such a procedure is necessary in order to remove the effects
of global rotation and translation of the molecule.

\begin{figure}[h]
\begin{tabular}{c}
\textbf{(a)} \\
\includegraphics[width=8.6cm]{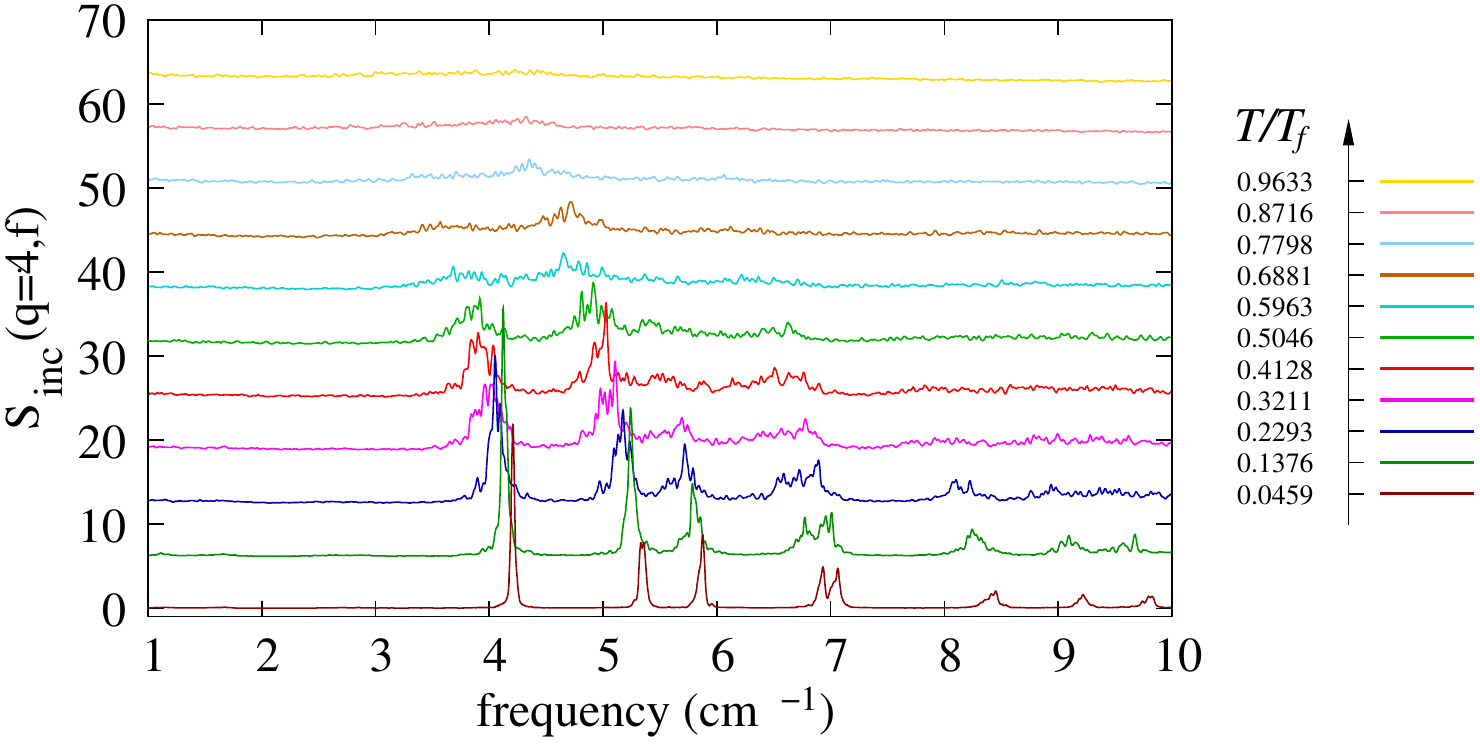} \\
\textbf{(b)} \\
\includegraphics[width=8.6cm]{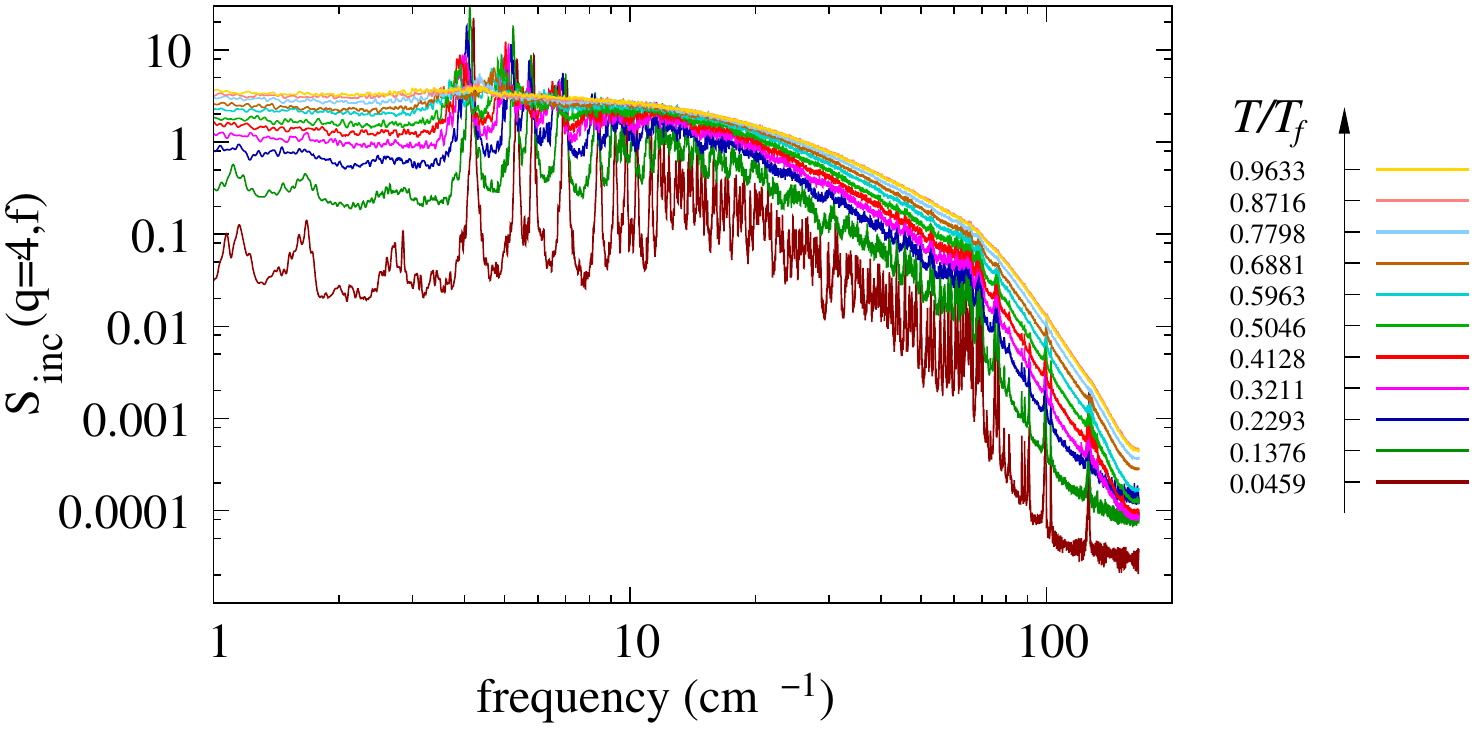}
\end{tabular}
\caption{(Color online) Incoherent structure factor $S_{inc}(\mathbf{q}=4$
  \AA$^{-1},\omega)$ as a function of temperature for protein G
  (Nos\'e-Hoover thermostat). 
  The unit of time has been converted to
  absolute units using the approximate conversion factor
  $1\;$t.u. $=0.1\;$ps.
 Panel a) shows a magnification of the structure factor in the
 low frequency range. The
  structure factors on this panel have been shifted with an
  offset to avoid the overlap of curves at different temperatures.
  The different curves, from bottom to top, correspond to the
  temperatures $T/T_f$ listed on the side of each panel.} 
\label{fig425}
\end{figure}

\smallskip
Figure \ref{fig425} shows the frequency dependence
of the incoherent structure factor $S(q,\omega)$ for a fixed wave vector
$q=4$ \AA$^{-1}$ for a simulation with the Nos\'e-Hoover
thermostat. On Fig.~\ref{fig425}-a, the evolution
of the low frequency range of the structure is shown for a range of
temperatures including the supposed dynamic transition region $T/T_f =
[0.4,0.5]$. At low temperatures
up to $T/T_f \approx 0.51$, individual modes are clearly distinguishable and
become broadened as temperatures increases. The slowest mode, located
around $4\;{\mathrm{cm}}^{-1}$ is also the highest in amplitude. It has a time
constant of about $\tau=80$ in reduced units ($\approx 8\;$ps). These
well-defined lines are observed to be shifting towards lower frequencies
with increasing temperature, similar to the phonon frequency shifts that
are frequently observed in crystalline solids. As we show in the
following section, the location of these lines can be calculated from a
harmonic approximation associated to a single potential energy
minimum. Therefore, the shift in frequency and the appearance of
additional modes can be seen as a signature of increasingly anharmonic
dynamics involving several minima associated to different conformational
substates.  
If, instead of the Nos\'e-Hoover thermostat, we consider the 
results obtained
with Langevin dynamics and a friction constant $\gamma=0.01$, 
the stronger coupling to the thermostat 
leads to low energy modes which are significantly broader than with
the Nos\'e-Hoover thermostat, so that
they can hardly be resolved anymore. However the location of the 
peaks in the spectra remains the same as the one shown on
Fig.~\ref{fig425}-b. Besides the larger damping, Langevin
calculations pose additional technical difficulties because 
Langevin dynamics does not preserve the total momentum of the system.
The center of mass of the protein diffuses on the time scale of the
trajectories. At low temperatures when fluctuations are small, the
alignment procedure can efficiently eliminate contributions from
diffusion as the center of mass is well defined for a rigid
structure. At high temperatures however, it cannot be excluded that the
alignment procedure adds spurious contributions to the structure factor
calculations as the fluctuations grow in amplitude and the structure
becomes flexible.

\bigskip
The analysis of the incoherent structure factor has shown that the low
temperature dynamics of the G\=o-model is dominated by harmonic
contributions. An increase of temperature leads to a broadening and a
shift of these modes until they eventually become continuously
distributed. However, for both strong and weak coupling to the
heat bath, no distinct change of behavior can be
detected within the temperature range $T/T_f = [0.4,0.5]$
in which Fig.~\ref{fig:deltar2} shows an apparent dynamical transition. 
Instead, the numerical results suggest a continuous
increase of anharmonic dynamics, and the absence of a dynamical
transition in this model, even though, in the range  $T/T_f = [0.4,0.5]$,
the peaks of the structure factor in the Nos\'e-Hoover
simulations broaden significantly. In the lowest temperature range the
structure factor does not show any contribution reminiscent of a
Boson peak.

\subsection{Structure factor from normal mode analysis in the one 
phonon  approximation}

A further analysis can be carried out to determine whether the low
temperature behavior of the protein model
shows a complex glassy behavior or simply the properties of an harmonic
network made of multiple bonds.
The picture of a rough energy landscape of a protein with many minima
separated by barriers of different height does not
exclude the possibility that, in the low temperature
range, the system behaves as if it were in thermal equilibrium
in a single minimum of this multidimensional  space.
This would be the case if the time scale to cross the energy barrier
separating this minimum from its neighbor basins were longer than the
observation time (both in numerical or real experiments). In this
case, it should be possible to describe the low temperature behavior of
the protein in terms of a set of normal modes. To determine if this is
true for the G\=o model that we study, one can compare the spectrum
obtained from thermalized numerical simulations at low temperature (low
temperature curves on Fig.~\ref{fig425}) with the calculation of the
structure factor in terms of phonon modes, in the spirit of the
study performed in ref.~\cite{goupil-lamy} for the analysis  of 
inelastic neutron scattering data of staphylococcal nuclease at $25\;$K on
an all-atom protein model.

\smallskip
The theoretical
basis for a quantitative comparison is an approximate expression of
the quantum-mechanical structure factor $S(\mathbf{q},\omega)$ in the
so-called \textit{one-phonon limit} which only accounts for single
quantum process in the scattering events assuming harmonic dynamics of
the nuclei. In this approximation, the incoherent structure factor can
be written as 
\begin{align}
S_{inc}(\mathbf{q},\omega)&=\sum_{i}\sum_{\lambda} b_{i}^2 e^{\hbar
  \omega_{\lambda}\beta/2} e^{-2 W_{i}(\mathbf{q})} \hbar
|\mathbf{q.}\mathbf{e}_{\lambda,i}|^2 \nonumber \\
 & \times \left(4 m_i \omega_{\lambda}
  \sinh(\beta\hbar\omega_{\lambda}/2) \right)^{-1}
\delta(\omega-\omega_{\lambda})\ . \label{nmsf}
\end{align} 
Here, the indices $i$ and $\lambda$ denote the atom and normal modes
indices respectively. $\mathbf{e}_{\lambda,i}$ is the subvector relating
to the coordinates of particle $i$ of the normal mode vector associated
to index $\lambda$. $W_{i}(\mathbf{q})$ denotes the Debye-Waller factor,
which in the quantum calculation of harmonic motion reads
\cite{goupil-lamy} 
\begin{eqnarray}
W_{i}(\mathbf{q})&=&\sum_{\lambda} \frac{\hbar
  |\mathbf{q.}\mathbf{e}_{\lambda,i}|^2}{ m_i \omega_{\lambda} }
\left[2n(\omega_{\lambda})+1 \right]\ \ \ , 
\end{eqnarray}
$n(\omega)$ being the Bose factor associated to the energy level
$\omega$.

For the calculations of the structure factor in the
G\=o-model within this approximation, we average
$S_{inc}(\mathbf{q},\omega)$ on a shell of $\mathbf{q}$-vectors by
transforming the Cartesian coordinate vector $(q_x,q_y,q_z)$ into
spherical coordinates
$\mathbf{q}=q\cdot
\left(sin(\theta)cos(\phi),sin(\theta)sin(\phi),cos(\theta)\right)$, 
and generate a grid with $N_q$ points for the interval $\phi=[0,2\pi]$,
and $N_q$ points for $\theta=[0,\pi]$. With this shell of vectors, we can
evaluate the isotropic average $S_{inc}(q,\omega)$. In equation
(\ref{nmsf}), $\hbar$ appears as a prefactor to the Debye-Waller factor
$W_i(\mathbf{q})$ in the exponentials and in the inverse hyperbolic
function. In order to evaluate the structure factor in reduced units of
the G\=o-model, we therefore need to estimate the order of $\hbar$ in a
similar way as we did for the energy scale (see Appendix) 
by comparing the fractions 
\begin{eqnarray}
\frac{\hbar \omega}{k_BT_f}&=&\frac{\hbar' \omega'}{(k_BT_f)'} \ \ \ ,
\end{eqnarray}
the non-primed variables denoting quantities in reduced units. In the
numerical evaluation of equation (\ref{nmsf}), we discretize the
spectrum of frequencies from the smallest eigenvalue to the largest mode
into $10000$ grid points to evaluate the $\delta$-function. We use
$N_q=225 $ vectors to average on a shell of modulus $|\mathbf{q}|=4$
\AA$^{-1}$. The summation runs over all eigenvectors except for the six
smallest frequencies which are numerically found to be close to zero,
and result from the invariance to overall translation and rotation of
the potential energy function.

In a first step, we use the coordinates of the global minimum of the
G\=o-model for protein G corresponding to the inherent structure with
index $\alpha_0$ to calculate the Hessian of the potential energy
function. The second
derivatives are calculated by numerically differentiating the
analytical first derivatives at the minimum. As discussed in the
Appendix, due to the presence of frustration in the potential,
the experimental 
structure does not correspond to the global minimum of the model. The
difference between the minimum and the experimental structure is however
small,
with  root-mean square deviation $0.16$ \AA \ and notable changes in
position occurring only for a small number of residues located in the
second turn. 

\begin{figure}[h]
\begin{tabular}{c}
\textbf{(a)} \\
\includegraphics[width=2.8in]{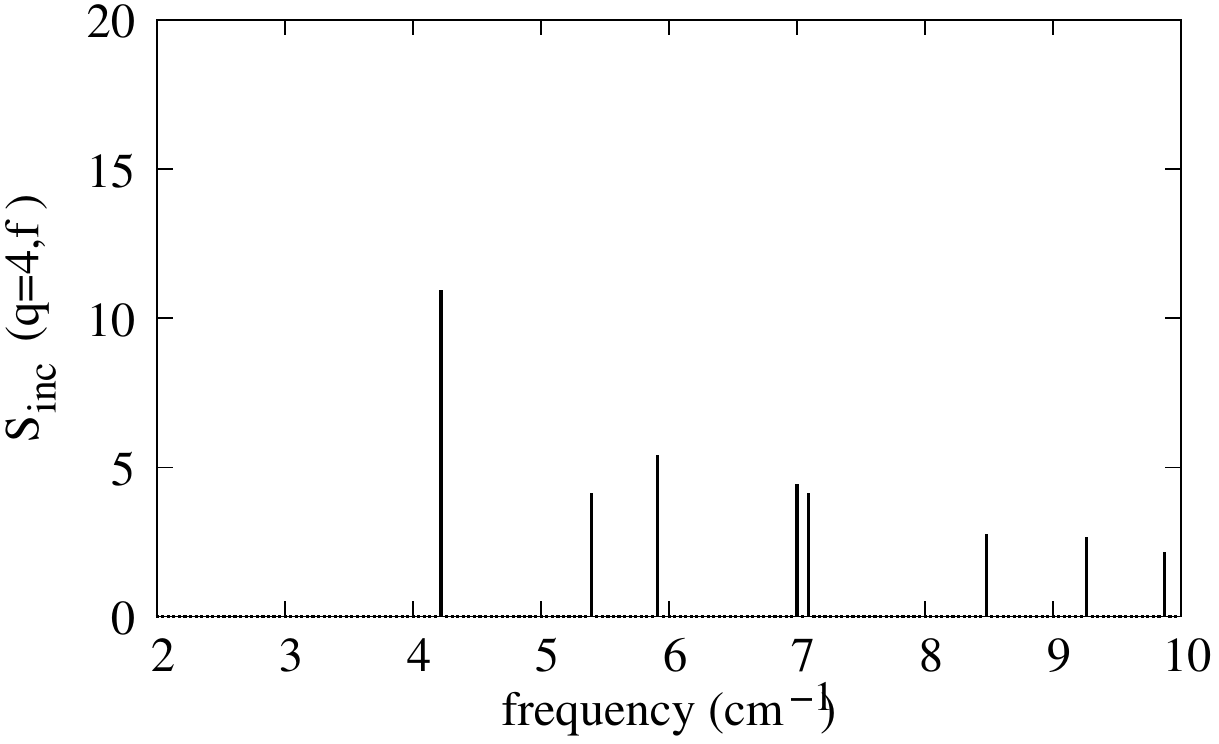} \\
\textbf{(b)} \\
\includegraphics[width=2.8in]{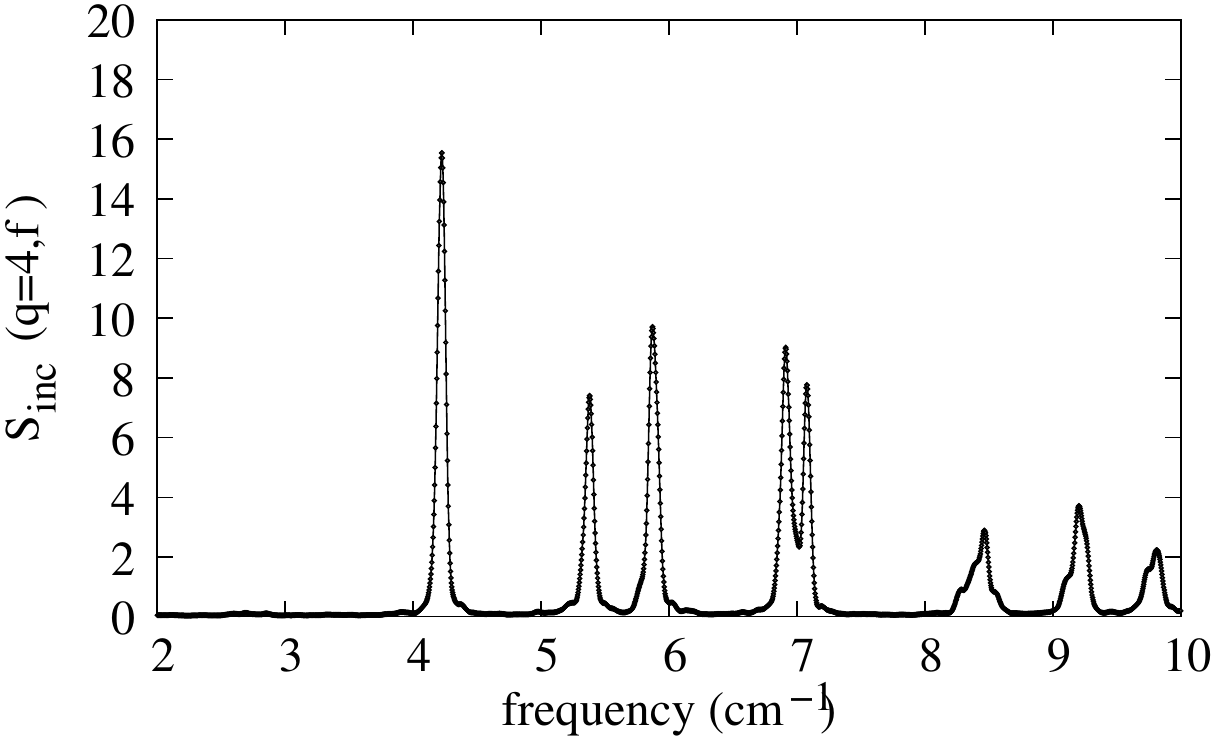}
\end{tabular}
\caption{a) Frequency dependence of the incoherent structure
  factor $S(q=4$ \AA$,\omega)$ ($T=0.0459 \, T_f$) calculated
  from normal modes in the one-phonon approximation. This figure only
  shows the lower frequency part of the spectrum. b)
  Incoherent structure factor $S(q=4$ \AA$,\omega)$ calculated from
  Nos\'e-Hoover constant temperature molecular dynamics at the same
  temperature.} 
\label{fig481}
\end{figure}

To estimate the
normal mode frequencies in absolute units, we use the conversion of the
time unit of $0.1\;$ps introduced in the Appendix. The
conversion into wave numbers, which is convenient for the comparison to
experimental data and to the results from all-atom calculations, is
achieved by noting that, from  $c k = f$, we can assign the conversion
$1\;{\mathrm{ps}}^{-1} \rightarrow  33.3\;{\mathrm{cm}}^{-1}$ 
and multiply the frequencies by
this scaling factor. Figure \ref{fig481}-a
 shows the results of the calculation  of the incoherent
structure factor $S(q=4$\AA$,\omega)$ in the one phonon approximation at
the temperature $T=0.0459 \, T_f$. Since in this approximation the normal
mode frequencies enter with a delta function into equation (\ref{nmsf}),
there is no line width associated to these modes unless the structure
factor is convoluted with an instrumental resolution function  or a
frictional model \cite{smith91}. Comparing to the structure factor
calculated from a molecular dynamics trajectory at the same temperature
(Fig.~\ref{fig481}-b), we find a
good correspondence of the location of the lines and their relative
amplitude with respect to each other. 

Therefore the analysis of the incoherent structure factor using a harmonic
approximation quantitatively confirms the dominant contribution of
harmonic motion at low temperatures. In particular, the motion at
very low temperatures occurs in a single energy well associated to one
conformational substate. To see how this behavior changes with
increasing temperature, in the next section we analyze the distribution of
inherent structures with temperature. 

\section{Inherent structure analysis in the dynamic 
transition region}  
\label{sec:is}

The freezing of the dynamics of a protein at temperatures below the
``dynamic transition'' is also described as a ``glass transition''. This
leads naturally to consider an energy landscape with many metastable
states, also called ``inherent states'' in the vocabulary of glass
transitions. In refs.~\cite{nakagawa2,hagmann1} we showed that the
thermodynamics of a protein can be well described in terms of its
inherent structure landscape, i.e.\ a reduced energy landscape which
does not describe the complete energy surface but only its minima. This
picture is valid at all temperatures, including around the folding
transition and above. For our present purpose of characterizing the low
temperature properties of a protein and probe its possible relation with a
glassy behavior, it is therefore useful to examine how the protein
explores its inherent structure landscape in the vicinity of the dynamic
transition. Here, we shall try to find how
the number of populated minima changes with temperature around the
transition region $T/T_f = [0.4,0.5]$ for the G\=o-model of protein G,
and which conformational changes can be associated to these inherent
structures.

\begin{figure}[h]
  \begin{tabular}[c]{c}
\includegraphics[width=7.6cm]{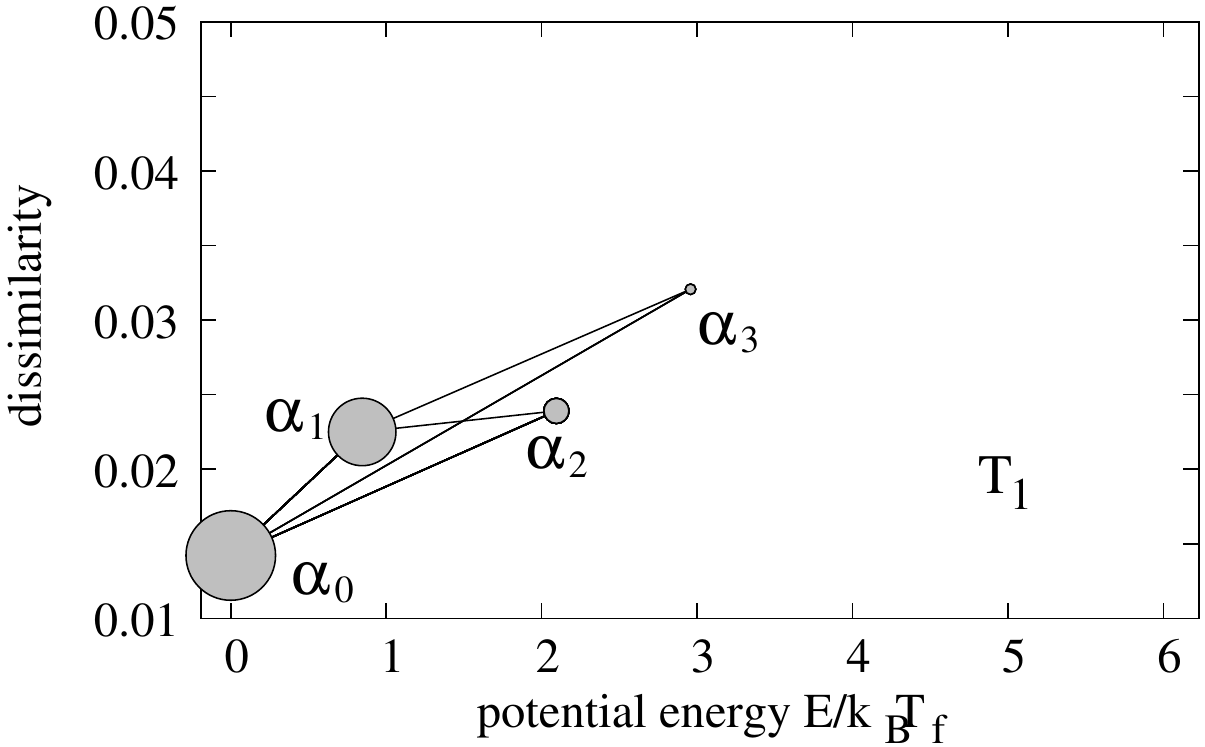} \\
\includegraphics[width=7.6cm]{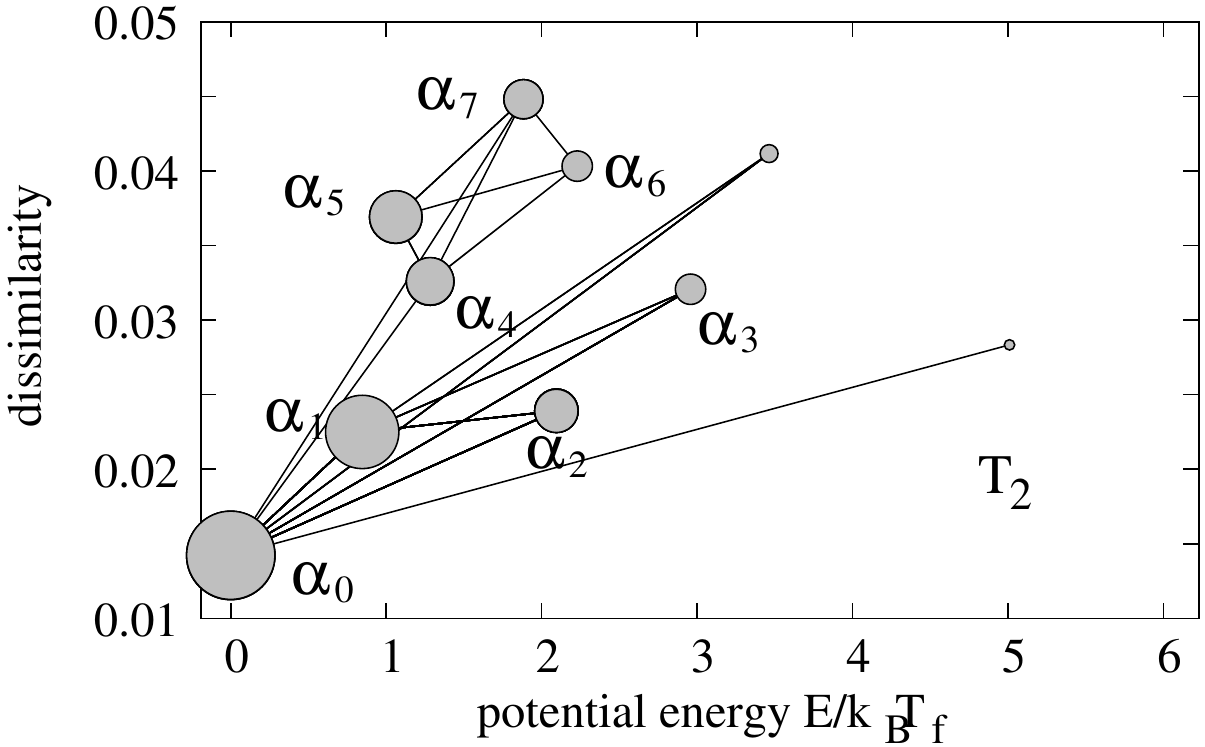} \\
\includegraphics[width=7.6cm]{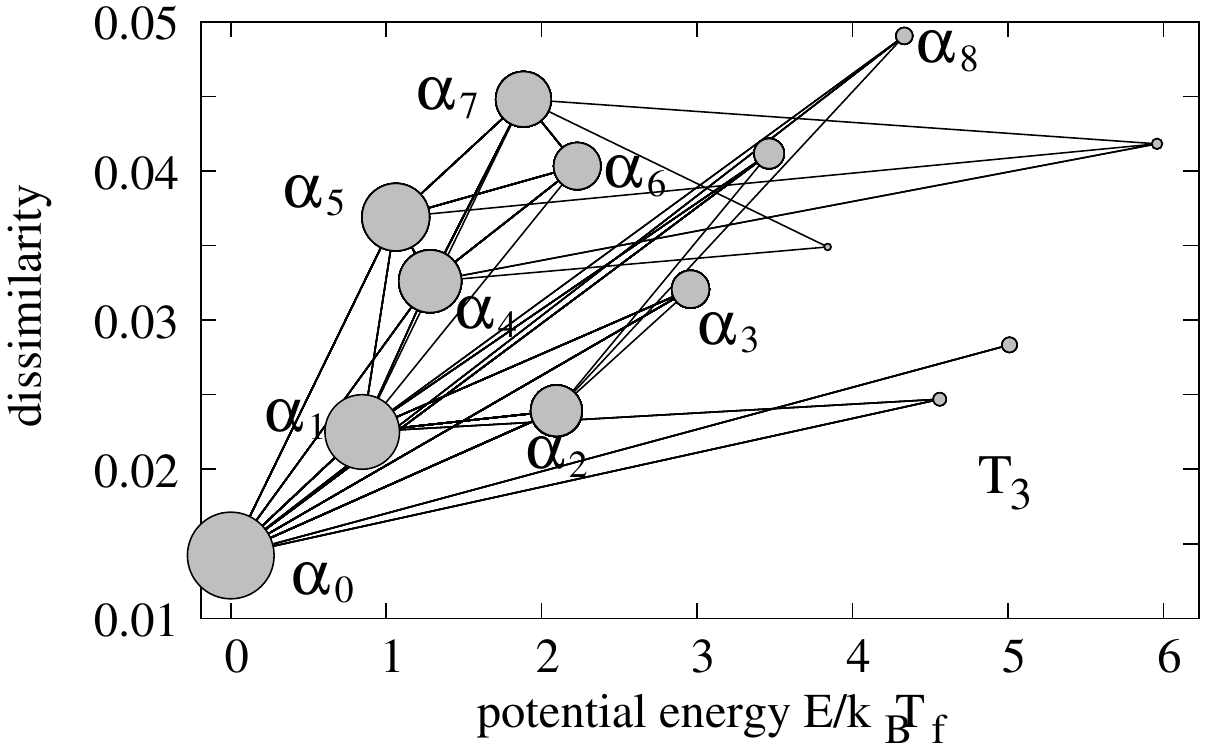} \\
  \end{tabular}
\caption{Inherent
  structure population of the G\=o-model for protein G at temperatures
  $T_1,T_2,T_3$ (from top to bottom) 
  and their associated structural dissimilarity. Lines are
  drawn between states that are connected within a MD trajectory. The
  width of the circles is proportional to $1/2\ log(w)$ where $w$ is the
  total number of occurrences of a given minimum.} 
\label{figihs}
\end{figure}

For three selected temperatures $T_1=0.275\ T_f$, $T_2=0.39\ T_f$,
$T_3=0.482\ T_f$ shown on Fig.~\ref{fig:deltar2},  
we generated $10$ trajectories from independent
equilibrated initial conditions for $2\cdot 10^7$ reduced time units
using the Nos\'e-Hoover thermostat. Along each trajectory, a
minimization was performed every $2\cdot 10^4$ time units such as to
yield $N_m=20000$ minima for each temperature point. In the
classification of these minima and their graphical representation, we
only keep those minima which have been visited at least $2$ times within
the $N_m$ minima, which lead to discard less than $10$ events
from the total number of counts. Most of the 
counts are concentrated on a small number of inherent structures. In
figure \ref{figihs}, we show the relative populations of the inherent
structures on a two-dimensional subspace spanned by the inherent
structure energy and the structural difference with the
experimental structure measured by the dissimilarity factor
\cite{nakagawa1,dissym}. The
radius of the circles centered at the location of the minima on this
plane is set proportional to $1/2\  \log(w)$ where $w$ is the absolute
number of counts of this minimum along the trajectories. This definition
is necessary to allow the graphical representation on the plane,
however, it may visually mask that linear differences in the radii
translate into exponential differences of the frequency of visit of the
minimum. As an example, the minima $\alpha_0,\alpha_1,\alpha_2,\alpha_3$
have the occupation probabilities $p(\alpha_0)=w(\alpha_0)/N_m\approx\
92\%$, $p(\alpha_1)\approx \ 8\%$, $p(\alpha_2)\approx\ 0.1\%$ and
$p(\alpha_3)\approx\ 0.02\%$ at $T=T_1$. 

\begin{figure}[h]
\centering
\includegraphics[width=8.0cm]{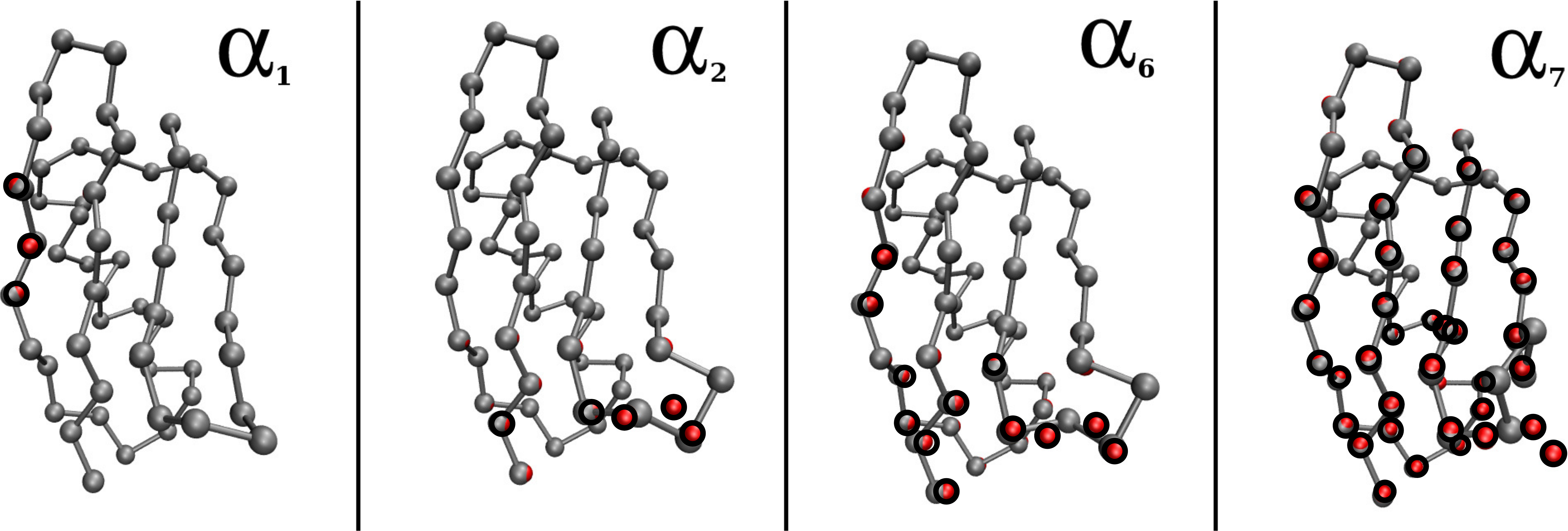}
\caption{(Color online) Shapes of
  inherent structures $\alpha_1$, $\alpha_2$, $\alpha_6$, $\alpha_7$. 
  The reference coordinates of the global minimum
  $\alpha_0$ are shown by red balls surrounded by a thick black line. 
 The  coordinates of the global minimum
  are invisible for residues which overlap with the inherent structure
  coordinates.} 
\label{figihs2}
\end{figure}

From figure \ref{figihs}, we notice that already at $T_1$ more than one
minimum is populated though the global minimum $\alpha_0$ is
dominant. In these figures, lines are drawn between minima that are
connected along the trajectory, i.e. that form a sequence of events. It
should however be noted that since the sampling frequency is low, it
cannot be excluded that an intermediate corresponding to an additional
connection line is skipped. Connections between all minima may therefore
exist even though they did not appear in the sequences observed in this
study. Moving to higher temperatures $T_2$ and $T_3$, a larger number of
minima which are both higher in energy and structural dissimilarity
appear. As the temperature rises, their population numbers become more
important, as can be seen e.g. by inspection of the radii of the block
$\alpha_4$-$\alpha_7$ on figure \ref{figihs}. To
obtain a physical picture of the conformational changes associated with
these minima, it is useful to align their coordinates onto the
coordinates of the global minimum. The results of such an alignment are
shown in figure \ref{figihs2}. In this figure, the coordinates of the
effective G\=o-model particles located at the positions of the
$C_{\alpha}$-atoms for each amino-acid are drawn in red color. One
notices that  the conformational changes associated to
$\alpha_1$-$\alpha_3$ are small. 
It is interesting to notice that these small changes
already appear in the range of temperatures where the rise in
fluctuations seems to grow still linearly with the temperature. The next
higher minima involve in particular a reorientation of a turn within the
$\beta$-sheets of  a protein. The temperature range at which these
minima start to be populated coincides with the transition region
revealed by the mean-distance displacement $\langle \Delta r^2\rangle$,
suggesting that the anharmonic motion required to make transitions
between the basins of these minima is at its origin. 

\smallskip
We again observe that the dynamic transition region does not exhibit any
particular change of behavior that could deserve the name of ``transition'',
but rather a gradual evolution which gets noticeable in the range
$T/T_f = [0.4,0.5]$. In the next section we use a
non-equilibrium approach to reveal whether the dynamics below the
transition range can be characterized as ''glassy'' or  not.

\section{Test of the fluctuation-dissipation theorem (FDT) - a
  non-equilibrium approach}
\label{sec:fdt}

An alternative approach to study the low temperature transition, for which
equilibrium simulations take a significant amount of computer time,
consists in the test of the response of the protein to external
perturbations. Rather than waiting a long time to see rare fluctuations
dominating the average fluctuation at low temperatures, the system is
driven out of equilibrium on purpose to either observe the relaxation
back to equilibrium and its associated structural changes, or the
response to a continuous perturbation to be compared to fluctuations at
equilibrium.

The fluctuation-dissipation theorem (FDT) relates
the response to small perturbations and the correlations of fluctuations
at thermal equilibrium for a given system. In the past years, the
theorem and its extensions have become a useful tool to characterize
glassy dynamics in a large variety of complex systems
\cite{crisanti}. For glasses below the glass transition temperature, the
equilibrium relaxation time scales are very large so that thermal
equilibrium is out of reach \cite{berthier}. Consequently, the FDT
cannot be expected to hold in these situations, and the response
functions and correlation functions in principle provide distinct
information. In this section, we test the FDT for the G\=o-model of
protein G at various temperatures to see whether a signature of glassy
dynamics is present in the system. To this aim, we first recall the
basic definitions and notations for the theorem.

In our studies we start from a given initial condition and put the
system in contact with a thermostat during a waiting time $t_w$. 
The end of the waiting time is selected as the origin of time ($t=0$)
for our investigation. If
$t_w$ is large enough (strictly speaking $t_w \to \infty$) the system is
in equilibrium at $t=0$.
We denote the
Hamiltonian of the unperturbed system $H_0$, which under a small
linear perturbation of the order $\epsilon(t)$ acting on an
observable  $B(t)$  becomes 
\begin{eqnarray}
H&=&H_0 - \epsilon(t) B(t) \ \ \ ,
\end{eqnarray}
where for $\epsilon=0$ we recover the unperturbed system. For any
observable $A(t)$, we accordingly define the two ensemble averages
$\langle A(t) \rangle_0^{t_w}$ and  $\langle A(t) \rangle_{\epsilon}^{t_w}$ where
the index references the average with respect to the
unperturbed/perturbed system respectively and the
exponent $t_w$ indicates how long the system was equilibrated
before the start of the investigation.  The correlation function
in the unperturbed system relating the observables $A(t)$, $B(t')$ at
two instances of time $t,t'$ is defined by 
\begin{eqnarray}
C_{AB}(t,t')&=&\langle A(t) B(t')\rangle_0^{t_w} - 
\langle A(t) \rangle_0^{t_w} \;\cdot \langle B(t')\rangle_0^{t_w} \ \ \ .
\label{eq:defcorrel}
\end{eqnarray}
The susceptibility $\chi_{AB}(t)$, which measures the time-integrated
response of the of the observable $A(t)$ at the instant $t$ to the
perturbation $\epsilon(t')$ at the instant $t'$, reads 
\begin{eqnarray}
\chi_{AB}(t)&=&\int_{t_0}^{t}dt'\ \frac{\delta \langle A(t)
  \rangle_{\epsilon}^{t_w}}{\delta \epsilon(t')} \ \; . 
\end{eqnarray}
The index $B$ in the susceptibility indicates that the
response is measured with respect to the perturbation arising from the
application of $B(t)$, and the lower bound $t_0$ of the
integral indicates the instant of time 
at which the perturbation has been switched on.

The integrated form of the FDT states
that the correlations and the integrated response are proportional and
related by the system temperature at equilibrium 
\begin{align}
& \chi_{AB}(t) =\frac{1}{k_BT} \; \Delta C \quad \mathrm{with}
\nonumber \\
& \Delta C = \left( C_{AB}(t,t)-C_{AB}(t,0) \right) \ \
. \label{fldis} 
\end{align}
In the linear response regime for a sufficiently small and constant
field $\epsilon$, the susceptibility can be approximated as 
\begin{eqnarray}
\chi_{AB}(t)&\approx&\frac{\langle A (t) \rangle_{\epsilon}^{t_w} - \langle A
  (t) \rangle_{0}^{t_w}}{\epsilon} 
\label{eq:chilin}
\end{eqnarray}  
such that in practice, verifying the FDT accounts for the comparison of
observables on both perturbed and unperturbed trajectories.\\{\
}\\The basic steps for a numerical experiment aiming to verify the FDT
can be summarized as follows:\\{\ }
\begin{itemize}
\item  Initialize two identical systems 1 and 2; 1 to be simulated with and 2
without perturbation.
\item  Equilibrate both systems without perturbation during $t_w$.
\item At time $t_0$ (in practice $t_0=0$, i.e.\ immediately after the
end of the equilibration period) 
switch on the perturbation for system 1 and acquire
  data for both systems for a finite time $t_{\mathrm{FDT}}$.
\item Repeat the calculation over a large number of initial conditions
  to yield the ensemble averages $\langle\cdot\rangle_0^{t_w}$ and
  $\langle\cdot\rangle_{\epsilon}^{t_w}$; combine the data according to
  equation (\ref{fldis}).
\end{itemize}

The protocol may be modified to include an external perturbation which
break the translational invariance in time. For instance
the initial state can result from a quench from a high to a low
temperature. Then the system is only equilibrated for a short time
$t_w$ before the perturbation  in the Hamiltonian is switched on. 
In this case the
distribution of the realizations of the initial conditions is {\em not the
equilibrium distribution} so that the correlation function defined by
Eq.~(\ref{eq:defcorrel}) depends on the two times $t$ and $t'$ and not
only on their difference.

\subsection{Simulation at constant temperature}

This case corresponds to $t_w \to \infty$. In our calculations we
start from an initial condition which as been thermalized for at least
$5000\;$ time units. 
The first step is to make an appropriate choice for the perturbative potential
$\epsilon(t) B(t)$. An earlier application of the FDT to a protein model
\cite{hayashi} has used the perturbative term
\begin{equation}
  \label{eq:pothayashi}
  \epsilon(t) B(t) =  \epsilon \sum_{i=1}^N \cos(k \; y_i) \; ,
\end{equation}
where $k$ is a scalar, $y_i$ the $y$ coordinate of amino-acid $i$ and
$\epsilon \not= 0$ for $t>t_0$ a constant.
This perturbation is invariant neither 
by a translation of the system nor by its
rotation. Although this does not invalidate the FDT, this choice poses
some problems for the accuracy of the calculations because, even in the
absence of internal dynamics of the protein, the perturbation varies as
the molecules diffuse in space or rotate. To avoid this difficulty we
selected the perturbation
\begin{equation}
  \label{eq:potpert}
  W := -  \epsilon B(t) 
  = -  \epsilon \sum_{i=1, i\not= 28}^N \cos(k \; r_{i,28}) \; ,
\end{equation}
where $r_{i,28}$ is the distance between amino-acid $i$ and the
amino-acid $28$ which has been chosen as a reference point within the
protein because it is located near the middle of the amino-acid
chain. Such a potential only depends on the internal state of the
molecule, while it remains unaffected by its position in space.
To test the FDT for the G\=o-model of protein G using Eqs. (\ref{fldis})
and (\ref{eq:chilin}), we add this potential $W$ to the potential
energy $V$ of the model
and we select $A(t) = B(t)$. The thermal fluctuations
are described with the same Langevin dynamics as previously. 
We switch on the perturbation for the equilibrated protein
model and record $50000$ to $400000$ trajectories (depending on the
value
of $\epsilon$) of duration $2000$ time units
for temperatures in the range $T/T_f = [0.275,0.826]$ covering both
the low temperature domain and the approach of the folding transition of the
protein. The perturbation prefactors chosen in this first set of
simulations were $\epsilon=0.05$ and $\epsilon = 0.005$, 
and the wave number of the cosine-term was $k=2 \pi / 10$.

\begin{figure}[h]
  \begin{tabular}[c]{c}
\includegraphics[width=7.6cm]{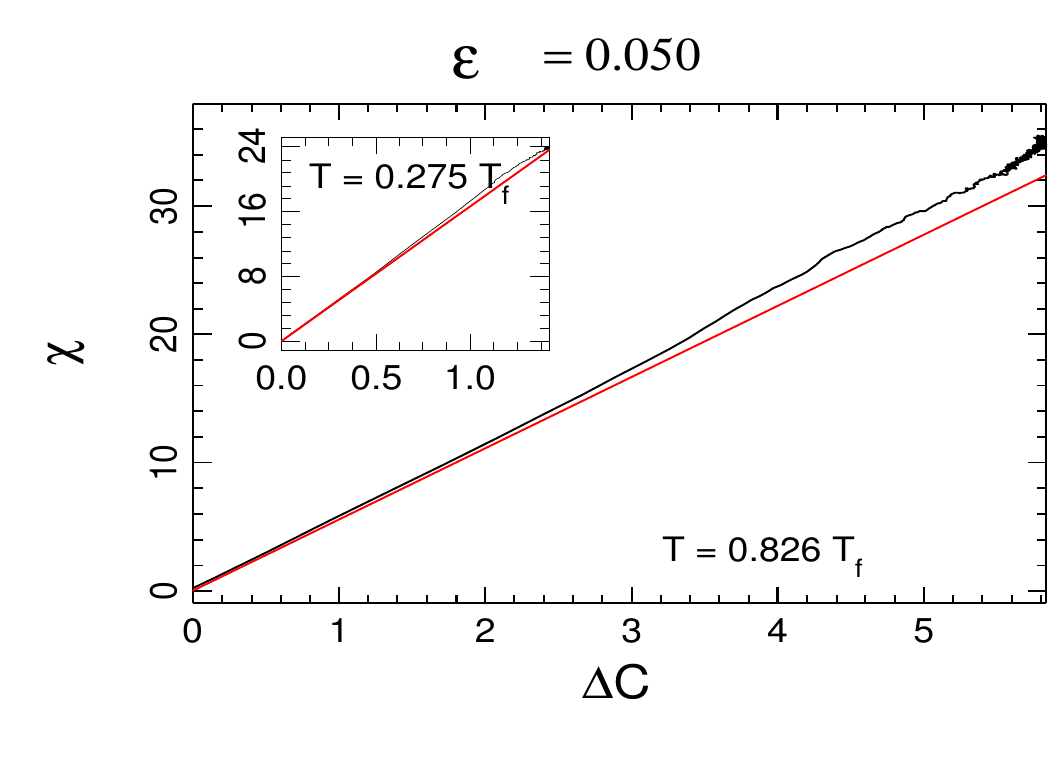} \\
\includegraphics[width=7.6cm]{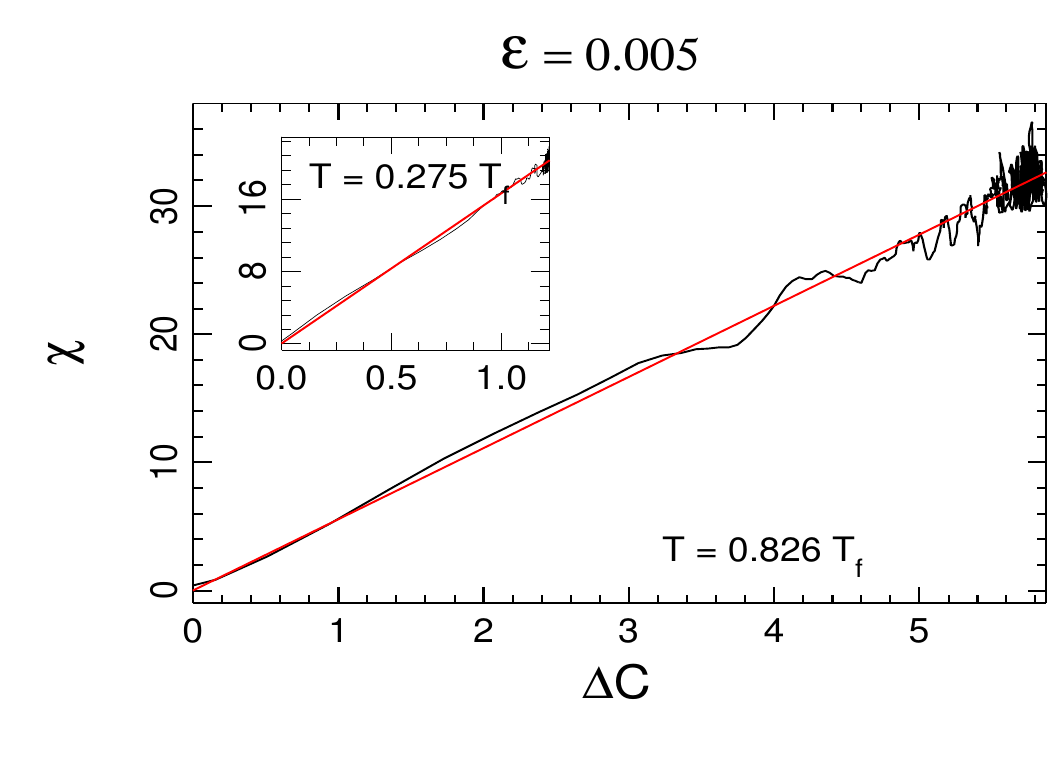}
  \end{tabular}
  \caption{(Color online) 
  Variation of
  $\chi$ versus $C$ for  
  $\epsilon = 0.05$ and $\epsilon = 0.005$ at the equilibrium 
  temperature $T = 0.826 \,  T_f$. 
The
  insets show $\chi$ versus $\Delta C$ at $T = 0.275 \,  T_f$. 
  The oblique (red)
  lines show the slope $1/k_BT$ that would be expected according to the
  fluctuation-dissipation theorem.
  The results presented on this figure have been obtained from 400000
  realizations.
}
  \label{fig:fdtequilibrium}
\end{figure}

Figure \ref{fig:fdtequilibrium} shows the evolution of the relation between the
susceptibility and the variation of the correlation function 
 $\Delta C = C_{AB}(t,t)-C_{AB}(t,0)$.
The straight lines represent the slopes
expected from the FDT. One notices that, for $\epsilon = 0.05$,
at $T = 0.826 \, T_f$, in the long term the value of 
$\chi / \Delta C$ stabilizes around a value which is away from
the expected value $1/k_B T$.
From a first glance, this result is reminiscent of 
the properties of a glass driven out of
equilibrium. In this context, the deviation from the slope expected from
the FDT is interpreted as the existence of an ''effective temperature''
for non-equilibrium systems. For the case studied here, 
finding an effective temperature would be surprising as the results are
obtained from measurements on a thermalized protein model, i.e.\ a system
in a state of thermal equilibrium. How is it then possible to explain
the apparent deviation from the FDT? 
The calculations performed with $\epsilon =
0.005$ give the clue because they show that the deviation 
appeared because the perturbation was too
large and outside of the linear response regime assumed to calculate the
susceptibility because for this lower value of $\epsilon$ the deviation
has vanished. If one computes the average value of the perturbation
energy $\langle W \rangle$ and compares it to the protein average energy
$\langle E(T) \rangle$, for the case shown on
Fig.~\ref{fig:fdtequilibrium}, $\epsilon = 0.05$, one finds
$\langle W \rangle / \langle E(T) \rangle = 1.3\cdot 10^{-2}$. This is small,
but, at temperatures which approach $T_f$ the protein is a highly
deformable object and even a small perturbation can bring it out of the
linear response regime. This shows up by the the a rise of $\chi$
versus time  for $\epsilon = 0.05$. At low temperatures the protein is more
rigid an therefore more resilient to perturbations. The insets on 
Fig.~\ref{fig:fdtequilibrium} show that for $\epsilon = 0.05$ the
calculations find that the fluctuation-dissipation 
relation at $T = 0.275 \, T_f$ is almost
perfectly verified although a very small deviation can still be
detected for this value of $\epsilon = 0.05$. Therefore  
a careful choice of parameters is
necessary to test the FDT under controlled conditions. In
particular, the perturbation needs to be carefully chosen to only
probe the internal dynamics and not to dominate them. 

\subsection{Simulation of quenching}

A typical signature of a glassy system is its aging after a
perturbation. In the context of the protein ``glass transition'', one
can therefore expect to detect a slow evolution of the system as a function
of the time after which it has been brought to the glassy state. This is
usually tested in quenching experiments, which can be investigated by a
sharp temperature drop in the numerical simulations. Our calculations
start from an equilibrium state at high temperature $T =1.40\,T_f$, 
which is abruptly cooled at a temperature $T_q$
below the temperature of the dynamical transition
studied in the previous sections. The model protein is then maintained
at this temperature $T_q$ by a Langevin thermostat. After a
waiting time $t_w$ we start recording the properties of the system over
a time interval  $t_{\mathrm{FDT}} = 25000\;$t.u. 
to probe the fluctuation dissipation
relation. In order to avoid nonlinear effects we use a small value of
$\epsilon = 0.005$. For such a weak perturbation, the response is weak
compared to thermal fluctuations and a large number of realizations
($50000$ or more) 
is necessary to achieve reliable statistical averages. To properly probe
the phase space of the model, these averages must be made over different
starting configurations before quenching. This is achieved by starting
the simulations from a given initial condition properly thermalized at
$T = 1.40\,T_f$ in a preliminary calculation. Then we run a
short simulation at this initial temperature, during which the
unfolded conformations change widely from a run to another with different random
forces because at high temperatures the fluctuations of the protein are
very large. The conformations reached after this short high-$T$
thermalization are the conformations which are then quenched to $T_q$,
for the FDT analysis.

\begin{figure}[h]
  \begin{tabular}[c]{c}
\includegraphics[width=7.6cm]{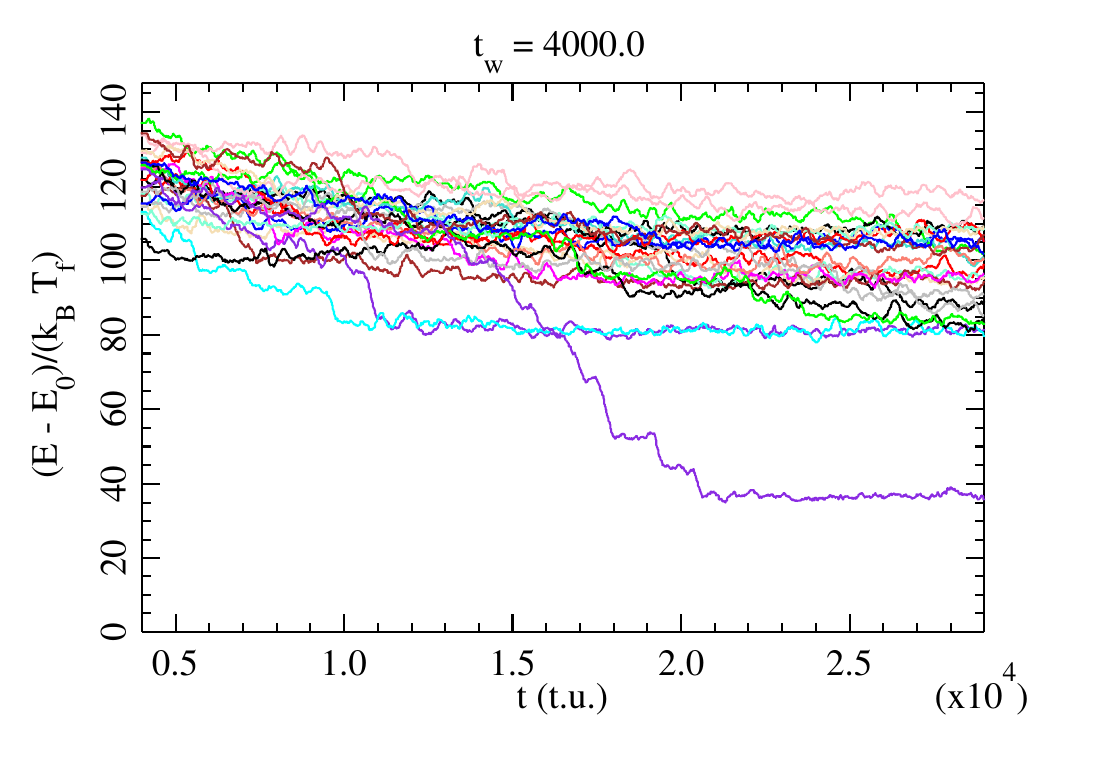} \\
\includegraphics[width=7.6cm]{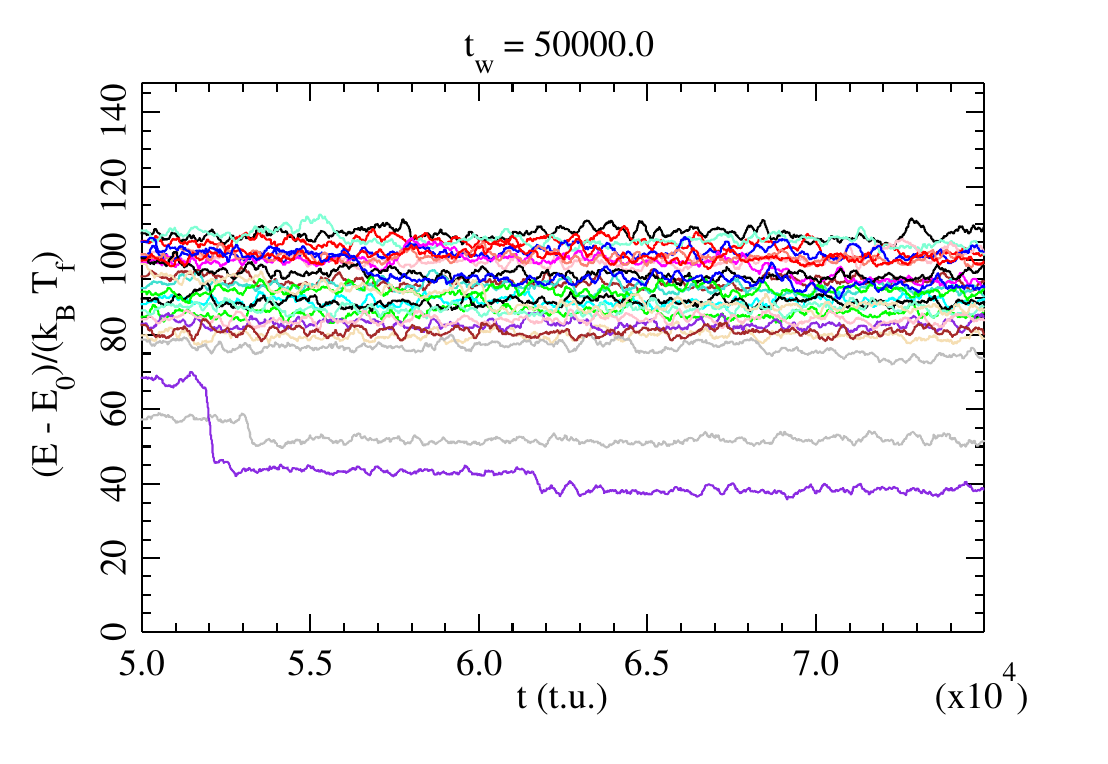} 
  \end{tabular}
\caption{(Color online)
Time evolution of the energy above the ground state, in units of
$k_B T_f$, after a  temperature jump from
  $T = 1.4\, T_f$ to $T  = 0.367\, T_f$  for two different waiting times
$t_w$, indicated in the title of each panel. The figure shows the
evolution of the energy for 15 realisations (corresponding to
the different colors) for the time $t_{\mathrm{FDT}} = 25000\;$t.u.
}
\label{fig:timequench}
\end{figure}

\begin{figure}[h]
  \begin{tabular}[c]{c}
\includegraphics[width=7.6cm]{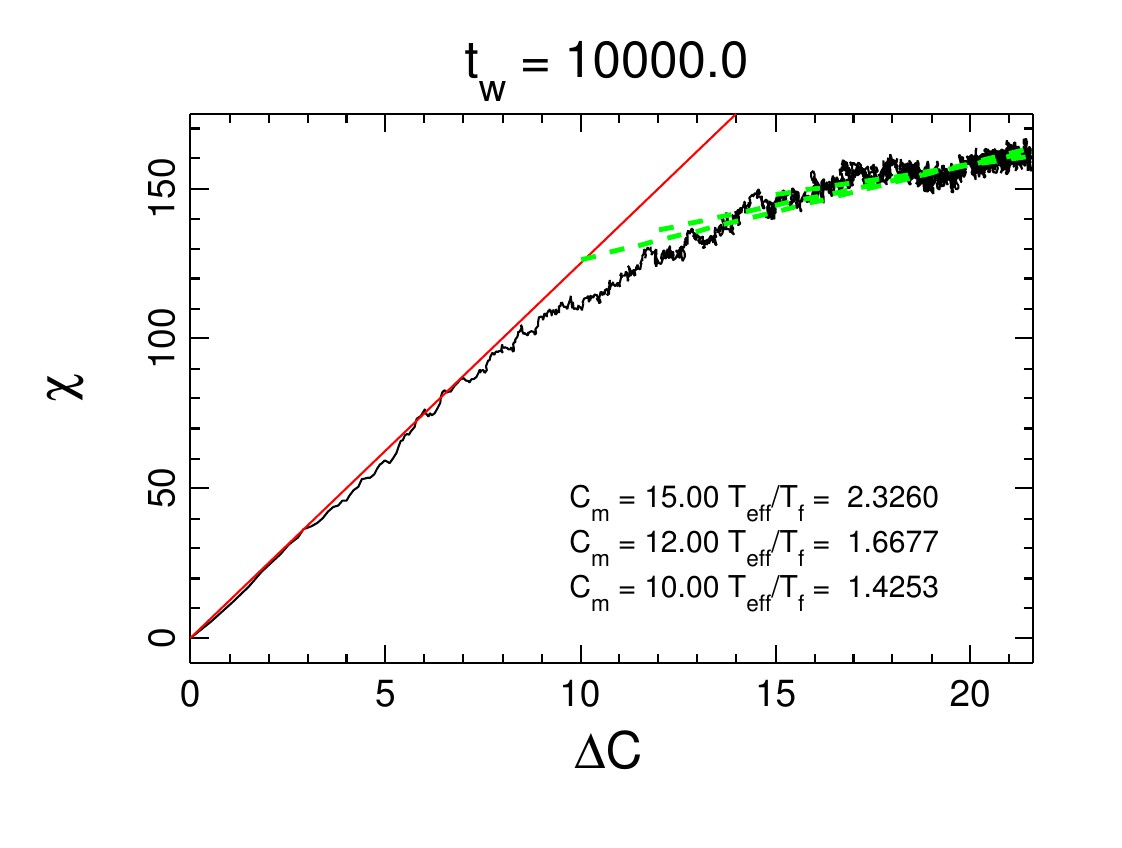} \\
\includegraphics[width=7.6cm]{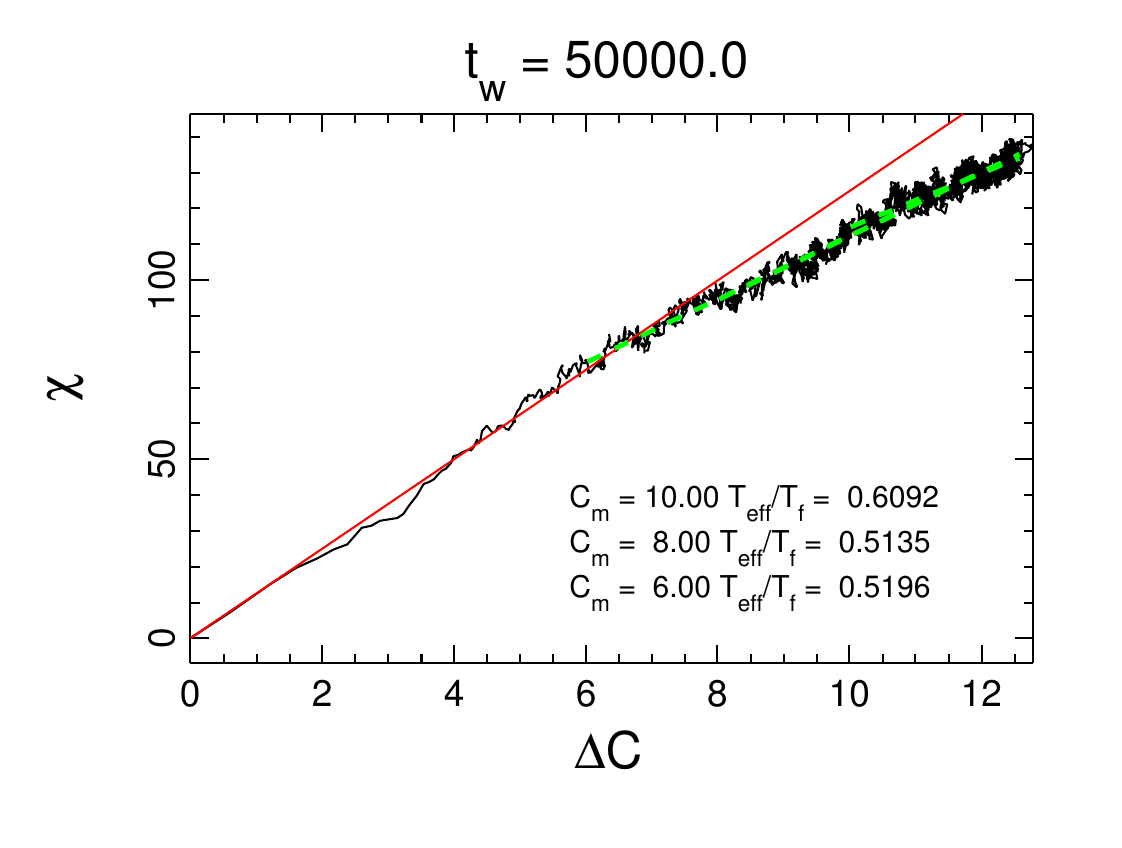} 
  \end{tabular}
\caption{(Color online) Test of the FDT for the temperature jump from
  $T = 1.4 \, T_f$ to $T = 0.367 \, T_f$ and different 
  waiting times $t_w$, indicated
in the title of each panel. The figures show the response function $\chi$
versus the variation $\Delta C$ of the correlation function. The
full straight lines (red) show the result of the FDT.
The results for $t_w = 10000\;$t.u. have been obtained by averaging over 115000
independent realizations, those for $t_w = 50000\;$t.u. have been obtained
with 92000 realizations. 
The dotted (green) lines show the fit of the
data for the domain $\Delta C > C_m$ where $C_m$ is the value above
which the curve deviates significantly from the line of slope $1/T$. The
inverse of the slope of these fits defines an effective temperature
$T_{\mathrm{eff}}$. As indicated in the legend of each figure, 
the value obtained for $T_{\mathrm{eff}}$ slightly depends on the choice of
$C_m$. By varying $C_m$ we can therefore estimate the
standard deviation on the value of $T_{\mathrm{eff}}$.
} 
\label{fig:fdquench}
\end{figure}

\smallskip
Typical results are shown on Figs.~\ref{fig:timequench} and \ref{fig:fdquench} 
for two values of $t_w$. The time evolution of the energy shows that, after the
waiting time $t_w$, even for the largest value $t_w = 50000\;$t.u. 
the model protein is still very far from equilibrium because its energy
is well above the ground state energy (chosen as the reference energy
$0$). This non-equilibrium situation sometimes leads
to rapid energy drops, generally accompanied by a decrease of
the dissimilarity with the native state, which superimpose to random
fluctuations which have to be expected for this system in contact with a
thermal bath. As expected the sharp variations of the conformations are
more noticeable for the shortest waiting time.
Figure \ref{fig:fdquench} shows that,
while for small values of $\Delta C = \left(
  C_{AB}(t,t)-C_{AB}(t,0) \right)$, which also correspond to shorter
times after we start to collect the data for the FDT test, the variation
of $\chi_{AB}(t)$ versus $\Delta C$ follows the curve given by the FDT
relation, then at larger $\Delta C$ the curve shows a significant
deviation from the slope $1/T$,
which defines an effective temperature $T_{\mathrm{eff}} > T$. The
effective temperature is larger for short waiting times after the quench
and decreases when $t_w$ increases. This should be expected because,
in the limit
$t_w \to \infty$ we should have $T_{\mathrm{eff}} \to T$ when the system
reaches equilibrium. 

\smallskip
It is not surprising to find a deviation from the FDT behavior after a
strong quench of the protein model because we put the system very
far from equilibrium. Therefore the observation of an effective
temperature that differs from the actual temperature of the system is
{\em not} a sufficient indication to conclude at the existence of a
glassy state of the protein model. What is important is the timescale at
which the system tends to equilibrium and how it depends on temperature.
To study this we have performed a systematic study of the variation of
$T_{\mathrm{eff}}(t_w;T)$ as a function of the waiting time $t_w$ and
temperature $T$, at the temperatures $T= 0.1875\,T_f$, $T=
0.2752\,T_f$, $T= 0.3670\,T_f$, $T= 0.4128\,T_f$ and
$T = 0.4817\,T_f$.
The temperature domain that we can study numerically is limited 
both from below and
from above. At the lowest temperatures the relaxation of the system is
very slow so that $t_w$ must be strongly increased. Moreover the speed at
which the protein model 
explores its phase space by moving from an inherent structure
to another becomes very low and statistically significant data cannot be
obtained without a large increase of $t_{\mathrm{FDT}}$. Running enough
calculations to get a good average on the realizations becomes
unpractical.
As discussed above, at high temperatures the protein becomes ``soft''
so that one quickly leaves the linearity domain of the FDT, unless the
applied perturbation becomes very small. But then the large thermal
fluctuations reduce the signal to noise ratio. Therefore the advantage
of faster relaxation times at high temperature is whipped out by the
need to make statistical averages over a much larger number of
realizations. However the temperature range over which one can get
statistically significant results overlaps the 
temperature $T \approx 0.45\,T_f$ above which the fluctuations of the
model appear to grow faster (Fig.~\ref{fig:deltar2}) so that one could
expect to observe a change in the properties of the system at this
temperature, if it existed.

\begin{figure}[h!]
  \centering
  \includegraphics[width=7.5cm]{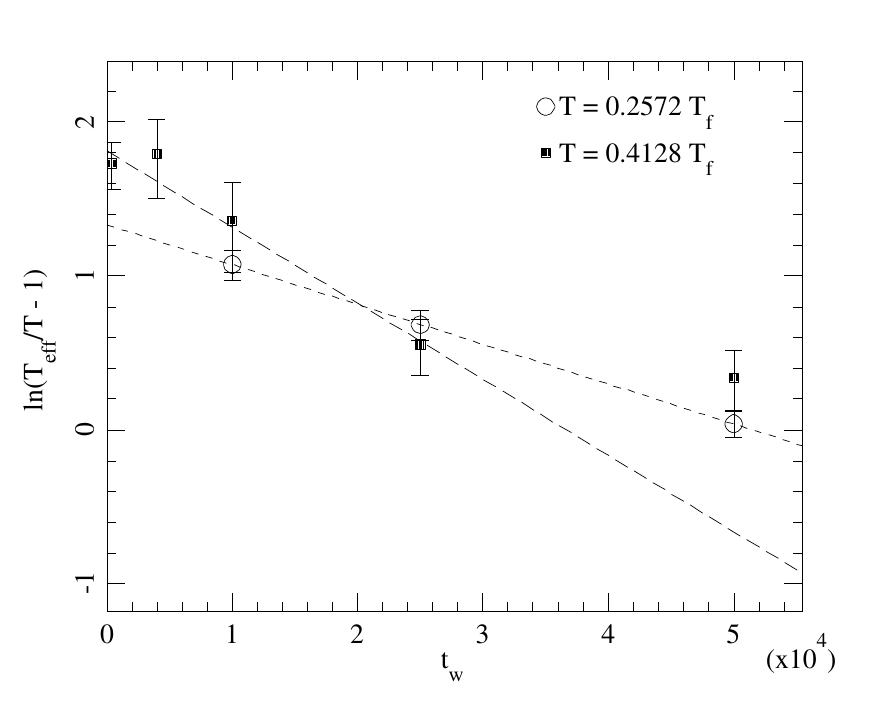}
  \caption{Variation of the effective temperature $T_{\mathrm{eff}}$ as
    a function of the waiting time $t_w$ at two temperatures,
    $T=0.2572\,T_f$ (open circles) and $T=0.4128\,T_f$ (closed
  squares). The figure shows $\ln \left( {T_{\mathrm{eff}}}/{T} - 1
  \right)$ versus $t_w$. The lines show a linear fit that takes into
  account the error bars on the determination of  $T_{\mathrm{eff}}$
  determined as explained in the caption of Fig.~\ref{fig:fdquench}.
 Such a fit determines $\tau(T)$ and its the standard deviation.
For
 $T=0.4128\,T_f$ the point corresponding to the largest value of
 $t_w$ is not included in the fit (see the discussion in the text).
}
  \label{fig:tefftw}
\end{figure}

\smallskip
At a given temperature $T$ we have defined a relaxation time $\tau(T)$
by assuming that the effective temperature relaxes exponentially towards
the actual temperature according to
\begin{equation}
  \label{eq:teffT}
  \left( \frac{T_{\mathrm{eff}}}{T} - 1 \right) \propto \exp  \left( -
    \frac{t_w}{\tau(T)} \right) \; .
\end{equation}
Figure \ref{fig:tefftw} shows that this assumption
is well verified by the numerical
calculations. It should however be noticed that, for the longest waiting
times, we may observe a large deviation from the exponential
decay, as shown in Fig.~\ref{fig:tefftw} for the results
at $T=0.4128\,T_f$. 
We attribute this to the limitations of our observations because,
when the system has sufficiently relaxed so that its effective
temperature approaches the actual temperature, all subsequent
relaxations become extremely slow and may exceed the observation time
$t_{\mathrm{FDT}} = 25000\;$t.u. so that the test of the FDT no longer
properly probes the phase space. Increasing $t_{\mathrm{FDT}}$ by an
order of magnitude might allow us to observe the relaxation further but
is beyond our computing possibilities as we have to study at least $50000$
realizations or more to achieve a reasonable accuracy.
A fit of the values of $ \left( {T_{\mathrm{eff}}}/{T} - 1 \right)$
versus $t_w$, which takes into account the statistical weight of each
point according to its standard deviation obtained from the
uncertainties on $T_{\mathrm{eff}}$ determines the relaxation time
$\tau(T)$ and its corresponding standard deviation.

\begin{figure}[h!]
  \centering
  \includegraphics[width=7.5cm]{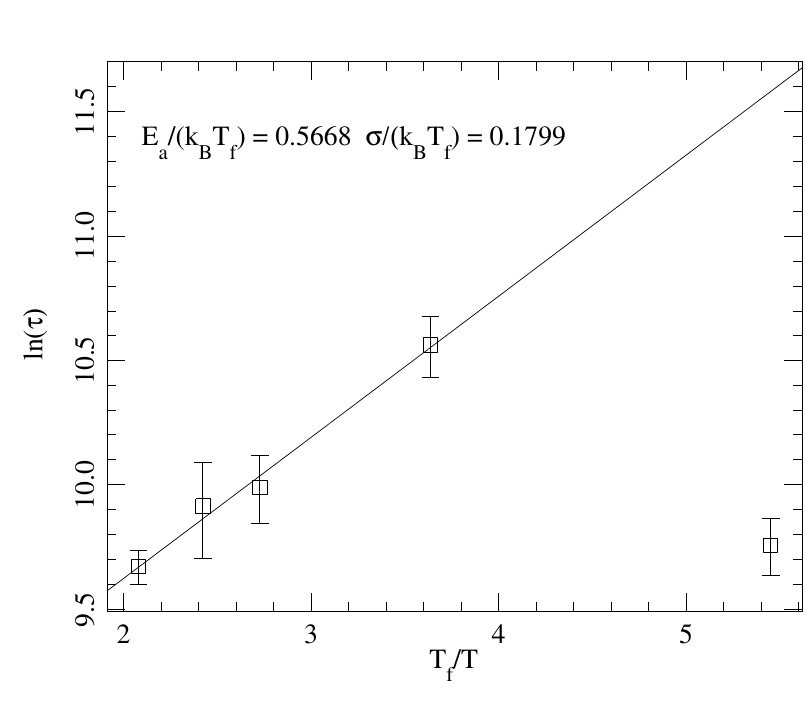}
  \caption{ Variation of the relaxation time $\tau(T)$ versus
    temperature. We plot $\ln(\tau)$ versus $T_f/T$. The line is a fit of
    the values which takes into account the uncertainties on the values
    of $\tau(T)$. }
  \label{fig:relaxtime}
\end{figure}

\smallskip
Figure \ref{fig:relaxtime} shows the variation of $\tau(T)$ with
temperature, in logarithmic scale, versus $T_f/T$. It shows that,
except for the value at the lowest temperature
$T =  0.1875\,T_f$, i.e. $T_f/T = 5.45$, within
the numerical errors evaluated at each stage of our calculation, the
relaxation time obeys a standard Arrhenius relation
\begin{equation}
  \label{eq:arrheniustau}
  \tau(T) = \tau_0 \exp\left( \frac{E_a}{T}\right)
\end{equation}
with an activation energy $E_a/(k_B T_f) = 0.5668 \pm 0.1799$. At the lowest
temperature, the relaxation temperature estimated from the
Arrhenius law is $\tau(T=0.1875\, T_f) \approx 100000\,$t.u. so that
calculations with $t_w \ge 100000$ as well as $t_{\mathrm{FDT}} \gg
100000$ would be necessary, which is unpractical. However 
the observed deviation from Arrhenius law at low $T$ cannot be attributed to a
low-temperature glass transition because such a transition 
would lead to a relaxation
time larger than predicted by the Arrhenius relation, while we observe
the opposite. In any case the  Arrhenius relation is well verified for a
temperature range which overlaps the temperature $T \approx
0.45\,T_{f}$, i.e.\ $T_f/T \approx 2.22$, 
above which dynamical simulations suggested a possible
increase of the fluctuations (Fig.~\ref{fig:deltar2}).
Therefore the relaxation of the protein model after a quench appears to
follow a standard activated process, with an activation energy 
of the order of $0.57 \;k_B T_f$, without any sign of a glassy behavior.

\bigskip
These observations can be compared with other studies of the
fluctuations of the same G\=o model of protein G
\cite{nakagawa2,NakagawaPRL}. Paper [\onlinecite{nakagawa2}] investigated
the fluctuation {\em in equilibrium} below the folding temperature
$T_f$. In these conditions, the numerical simulations of a protein which
is near its native state detect small, up and down, jumps of the
dissimilarity factor, which, in any case, stays very low ($d \approx 0.06$
for the equilibrium temperature $T=0.55 \, T_f$) but switches between
values that differ by $\approx 0.01$. In its equilibrium state the
protein may jump from an inherent structure to another but these
fluctuations are much slower than the one that we observed shorter after
a temperature quench because they occur on a time scale of the order of
$10^7\;$t.u. Their activation energy had been found to be $E_B =
6.2\,T_f$, i.e. much higher than the activation energy $E_a$ that
characterizes the relaxation of the effective temperature that we
measured. 
Those results are not in contradiction because they correspond
to fundamentally different phenomena. The non-equilibrium fluctuations
that we discuss in the present paper appear because the potential energy
surface has minima on the side of the ``funnel'' that leads to
the native state. Such minima, corresponding to protein structures which
are not fully folded, can temporarily trap the protein in intermediate
states.
However the lifetime of these high-energy minima is only of the order of
$10^5\;$t.u. i.e. they are
short lived compared to the residence time of the protein in an inherent
structure close to the native state. When the protein escapes from one
of these high-energy minima we observe an energy drop, as shown in
Fig.~\ref{fig:timequench}. 

\medskip
The study that we presented here is neither a study of the protein near
equilibrium, nor an investigation of the full folding process which also
occurs on much longer time scales (typically $10^7$--$10^8\;$t.u.) as
observed in Refs. [\onlinecite{nakagawa2}] and
[\onlinecite{NakagawaPRL}]. Therefore the activation
energy $E_a$ is also different from the energy barrier for folding.
For the same reason the effective temperature after the quench
$T_{\mathrm{eff}}$ should not be confused with the configurational
temperature $T_{\mathrm{cnf}}$ defined in Ref.~[\onlinecite{NakagawaPRL}]
which relates the entropy and energy of the inherent structures during
folding. $T_{\mathrm{cnf}}$ gives a global view of the phase space
explored during folding, and it evolves with a characteristic time of
$10^7$--$10^8\;$t.u. as the folding itself. Compared to these scales,
the FDT analysis that we presented here appears as a snapshot of the
strongly non-equilibrium state created after a fast quench. It offers a
new view of the time evolution of the protein model, which completes the
ones which had been presented earlier.

\section{Discussion}
\label{sec:discussion}

The starting point of our study has been the numerical observation of a
low temperature transition in a simplified protein model resembling the
experimentally observed dynamical transition in hydrated protein
samples. 
This suggested that, in spite of its simplicity, the frustrated G\=o
model could be used not only to study the folding of proteins but also
their dynamics in the low temperature range, opening the way for an
exploration of the glassy behavior of proteins.

\smallskip
We have therefore used different approaches to further characterize the
properties of the protein model in the low temperature
range. Thermalized molecular dynamics simulations have been used to
calculate the incoherent structure factor than one could expect to
observe for the protein. It shows peaks that broaden as temperature
increases, suggesting that the dynamics of the protein model is
dominated by harmonic or weakly anharmonic vibrations. This has been
confirmed by the calculation of the structure factor in the one-phonon
approximation. All the main features of the structure factor obtained by
simulations, such as the peak positions and
even the power law decay of the amplitude of the modes with frequency,
are well reproduced. In the low temperature range, the dynamics of the
protein appears to occur in a single energy well of its highly
multidimensional energy landscape. Of course this is no longer true
when the temperature increases. 

By analyzing the population of the
inherent structures, we have shown that, in the temperature range of the
dynamical transition, there is a continuous increase of the number of
states which are visited by the protein.  The transition
seems to be continuous, and it is likely that numerical observations
suggesting a sudden increase may have the origin in the limited
statistics due to finite time observation. Indeed, as shown for the
example of protein G, the conformational transitions become extremely
slow at low temperatures, such that the waiting time between the jumps
between conformations may exceed the numerical (or the experimental)
observation time. It is this breakdown of the ergodic hypothesis
together with the observation of non-exponential relaxation rates which
may have lead to the emergence of the terminology ''protein glass
transition'' in analogy to the phenomenology of glasses.

Non equilibrium studies allowed us to systematically probe a
possible glassy behavior by searching for violations of the
fluctuation--dissipation theorem. First we have shown that these
calculations must be carried out with care because apparent violations
are possible, even when the system is in equilibrium, due to
nonlinearities in the response. Except at very low temperatures they can
be observed even for perturbations as low as 1\% of the potential energy
of the system. Once this artifact is eliminated by the choice of a
sufficiently small perturbative potential, we have shown that, after a
sudden quench from an unfolded state to a very low temperature $T_q$,
{\em one   does observe} a violation of the FDT in the protein model,
analogous to what is found in glasses. The quenched protein is
characterized by an effective temperature $T_{\mathrm{eff}} > T_q$. But
{\em the relaxation of the model towards equilibrium}, deduced from the
evolution of the effective temperature $T_{\mathrm{eff}}$ as a function
of the waiting time after the quench, {\em follows a standard Arrhenius
behavior}, even when the temperature crosses the value $T \approx 0.45
T_f$ at which dynamical simulations appeared to show a change in the
amplitude of the fluctuations.

\bigskip
Although one cannot formally exclude that the results could be
different for
other protein structures or other simplified protein models,
this work concludes that a coarse-grain model such as
the G\=o model is too simple to describe the complex behavior of 
protein G and particularly its glass transition.
 Indeed such a model does
not include a real solvent, which plays an important role in
experiments. The thermostat used in the molecular dynamics simulations
only partially models the effect of the surrounding of the protein. The
apparent numerical transition previously observed for protein G may simply
be related to finite-time observations of the activation of structural
transitions which appears in a particularly long time scale for
proteins. This is an obvious limitation of molecular dynamics
calculations, but this could also sound as a warning to
experimentalists. Indeed experiments can access much longer time
scales. But they also deal with real systems which are much more complex
than the G\=o model. In these systems relaxations may become very long,
so that the experimental observation of a transition could actually
face the same limitations as the numerical simulations. 
Such a ''time window'' interpretation has
also been brought forward for the experimentally observed dynamical
transition, suggesting that the transition may in fact depend on the
energy-, and thus, on the time-resolution of the spectrometer
\cite{becker}.
In this respect,
as shown by our non-equilibrium studies to test the validity of the FDT,
such measurements, if they could be performed for a protein should tell
us a lot about the true nature of the ``glass transition'' of proteins.

\begin{acknowledgments}
  M.P. and J.G.H would like to thank Ibaraki University, where part of
  this work was made, for support. 
This work was supported by KAKENHI No 23540435 and No 25103002
Part of the numerical calculations
  have been performed with the facilities of the 
 P\^ole Scientifique de Mod\'elisation Num\'erique (PSMN) of ENS Lyon.
\end{acknowledgments}

\appendix*
\label{sec:appendix}
\section{Simulation and units}
\begin{figure}[h]
  \begin{tabular}[c]{c}
\includegraphics[width=7.5cm]{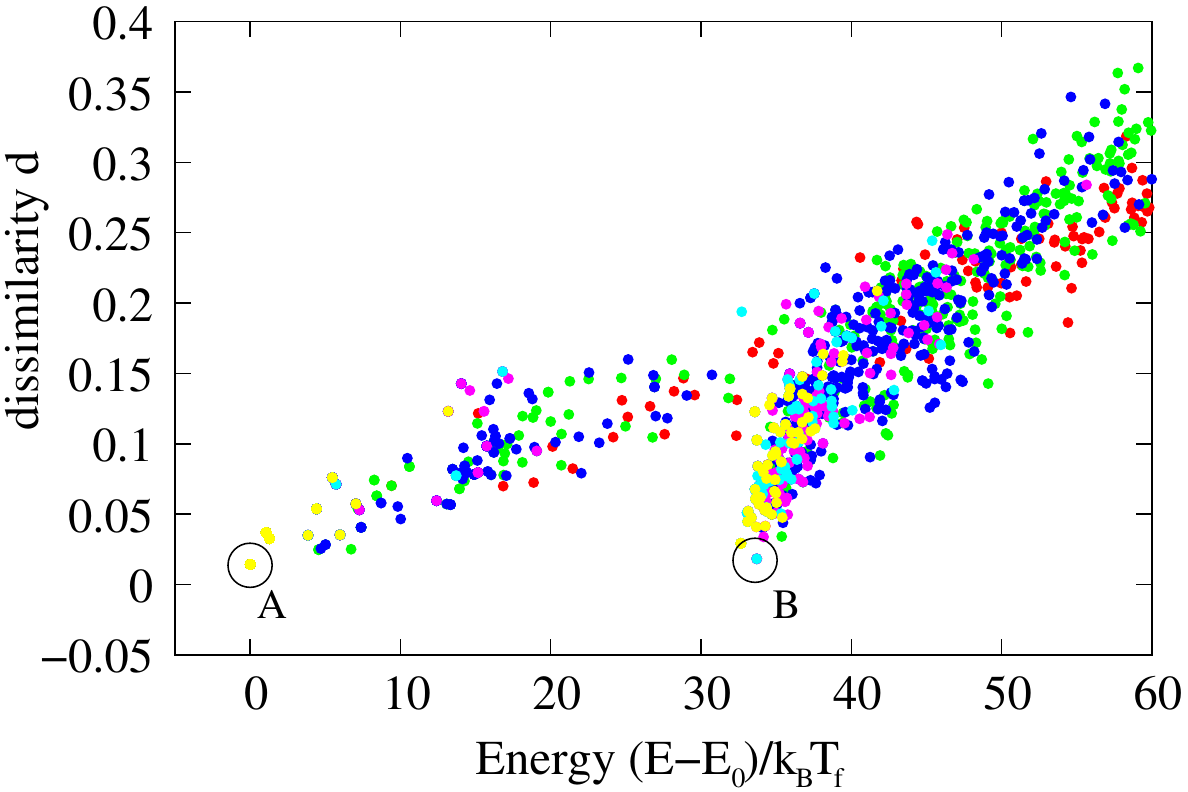}
  \end{tabular}
\caption{ Stable and metastable conformations of the
  protein $G$ model obtained by a non-equilibrium cooling protocol
  followed by energy  minimization (the various colors show minima obtained
  with different speed of cooling). 
  The horizontal axis shows the energy of each structure with reference to the
  global minimum,  and the vertical axis indicates its dissimilarity with the
  ground
  state \cite{nakagawa1,dissym}, lower values indicating more structural
  similarities between two conformations.
}
\label{fig:twostruct}
\end{figure}

\smallskip
The forcefield and the parametrization of the simplified G\=o-model are
presented in \cite{nakagawa2,nakagawa1}. In a standard G\=o model
the potential energy is written in such a way that the experimental
energy state is the minimal energy state. We use here a weakly
frustrated G\=o model for which the dihedral angle potential
does not assume a minimum in
the reference position defined 
by the experimentally resolved structure: it favors angles close to
$\pi/4$ and $3\pi/4$ irrespective of the secondary structure element
(helix, sheet, turn) the amino acid belongs to. 
This source of
additional ''frustration'' affects the dynamics and thermodynamics
of the model, leading to a more realistic representation
\cite{nakagawa1}. This feature
introduces additional complexity in the model
because, besides its ground state corresponding to the
experimental structure, the frustrated model exhibits another funnel for
folding,  which leads to a second structure which is almost a mirror
image of the ground state, but has a significantly higher energy 
(Fig.~\ref{fig:twostruct}).

To control temperature in the molecular dynamics simulations,
several types of numerical thermostats were used. Most of the
calculations use underdamped Langevin simulations
\cite{bbk,Honeycutt} with a time step 
$dt = 0.1\;$t.u. and friction constants in the range $\gamma = 0.01,
0.025$. The mass of all the residues is assumed to be equal to $m=10$. Some
calculations were also performed with the multi-thermostat Nose-Hoover
algorithm using the specifications defined in \cite{martyna}.

\smallskip

\noindent For simulation purposes, the variables in the G\=o-model are
chosen dimensionless (reduced units). Lengths are expressed in units of
$\tilde{l}=1$ representing \AA ngstr\"oms,
and the average mass of an amino-acid, $135\;$Da, has been
expressed as $10$ units of mass of the model. As the empirical
potentials are defined at the mesoscopic scale of the
amino-acids, values for the interaction constants in the effective
potentials cannot be easily estimated in absolute units. 
It is possible to estimate the energy scale of the
model by comparing the reduced folding temperature of the G\=o-model
with a realistic order of magnitude of the folding
temperature $T'_f$ in units of $K$ and setting 
\begin{eqnarray}
k'_B T'_f&=& \tilde{\epsilon} k_B T_f \ \ \ \ ,
\end{eqnarray}
where the variables on the left-hand side are given in SI units ($T'_f$
is the estimate of the folding temperature), and unprimed variables are
written in reduced units; $\tilde{\epsilon}$ is the required energy
scale in units of  $J$ to match between both. In our calculations
$k_B=1$ (meaning that reduced temperatures are expressed in reduced energy
units)  and $T_f=0.218$ is deduced from equilibrium simulations. 
Then a simple dimensional analysis gives the time unit of the model as
$
\tilde{t}=\sqrt{\frac{\tilde{m}\tilde{l}^2}{\tilde{\epsilon}}}\ .
$
One arrives at an estimated time
unit of $\tilde{t} \approx 0.1\;$ps. In the paper we refer
to $\tilde{t}\approx 0.1 \;$ps as the unit of time (t.u.) for the simulations
of the G\=o-model, keeping in mind that this can merely be seen as an
order of magnitude in view of the approximations that lead to this
number.

\bibliographystyle{apsrev}

\begin{thebibliography}{99}
\bibitem{dissym} 
The dissimilarity factor is a weighted distance map between two
conformations. Let $A$ and $B$ be two conformations of the protein
(for our purpose $B$ will be the native structure) and $a_{ij}$,
$b_{ij}$ 
the distance between residues $i$ and $j$ in these conformations. The
dissimilarity $d(A,B)$ between the two conformations is defined by
\begin{equation*}
  \label{eq:dissimilarity}
d(A,B) = \dfrac{\sum_{i=1}^{N_r} \sum_{j=i+2}^{N_r} 
|a_{ij}^{-p} - b_{ij}^{-p} | }%
{\frac{1}{2}\left( \sum_{i=1}^{N_r} \sum_{j=i+2}^{N_r} a_{ij}^{-p}
+ \sum_{i=1}^{N_r} \sum_{j=i+2}^{N_r}  b_{ij}^{-p} \right)}
\end{equation*}
where $a_{ij}$ and $b_{ij}$ are the distances between residues $i$ and
$j$ in the $A$ and $B$ conformations and $p$ an integer which
determines how much residues which are far apart contribute. We use here
$p=2$. In this article the reference conformation A is chosen as the NMR
structure, see \cite{nakagawa1} for details.
\end{thebibliography}

\end{document}